\documentstyle[12pt,graphicx,color,subfigure,amsmath,cite]{article}
\newcommand{\comment}[1]{}
\def\tanb{\tan\beta}

\def\thetab0{\theta_{B_0}}

\def\r2{\sqrt 2}
\def\beq{\begin{equation}}
\def\eeq{\end{equation}}
\def\beqn{\begin{eqnarray}}
\def\eeqn{\end{eqnarray}}
\def\sinW2{\sin^2\theta_W}

\def\mz2{M_{z}^2}
\def\c2b{\cos 2\beta}

\def\mz{M_Z}
\def\mhalf{m_{\frac{1}{2}}}

\def\rmGeV{\rm ~GeV}

\def\sec2w{sec^2\theta_W}
\def\gmin2{(g-2)_\mu}

\catcode`\@=11 % This allows one to modify PLAIN macros.

\def\lsim{\mathrel{\mathpalette\@versim<}}
\def\gsim{\mathrel{\mathpalette\@versim>}}
\def\@versim#1#2{\vcenter{\offinterlineskip
    \ialign{$\m@th#1\hfil##\hfil$\crcr#2\crcr\sim\crcr } }}

\def\mygraph#1#2{ \subfigure[]{
   \label{#1}
   \hspace*{-0.6in}                     
   \begin{minipage}[b]{0.5\textwidth}                       
   \centering
   \hspace*{4ex}
   \includegraphics[width=0.9\textwidth,height=0.7\textwidth]{#2}
   \vspace*{-7ex}
   \end{minipage}}
   \vspace*{-1ex}
}

\begin{document}
\begin{center}
  { \Large\bf 
Large evolution of the bilinear Higgs coupling parameter in SUSY models 
%with
%non-universal gaugino masses
and reduction of phase sensitivity
\\}
  \vglue 0.5cm
  Utpal Chattopadhyay$^{(a)}$, 
Debajyoti Choudhury$^{(b)}$ and Debottam Das$^{(a)}$\footnote{Emails: tpuc@iacs.res.in, 
debchou@physics.du.ac.in, tpdd@iacs.res.in}
    \vglue 0.2cm
    {\em $^{(a)}$Department of Theoretical Physics, Indian Association 
for the Cultivation of Science, Raja S.C. Mullick Road, Kolkata 700 032, India \\}
    {\em $^{(b)}$Department of Physics and Astrophysics, University of Delhi,  
Delhi 110 007, India \\} 
  \end{center}        
\begin{abstract}
The phases in a generic low-energy supersymmetric model are severely
constrained by the experimental upper bounds on the electric dipole
moments of the electron and the neutron.  Coupled with the requirement
of radiative electroweak symmetry breaking, this results in a large
degree of fine tuning of the phase parameters at the unification
scale. In supergravity type models, this corresponds to very highly
tuned values for the phases of the bilinear Higgs coupling parameter
$B$ and the universal trilinear coupling $A_0$.  We identify a
cancellation/enhancement mechanism associated with the renormalization
group evolution of $B$, which, in turn, reduces such fine-tuning quite
appreciably without taking recourse to very large masses for the
supersymmetric partners. We find a significant amount of reduction of
this fine-tuning in non-universal gaugino mass models that do not
introduce any new phases.
\end{abstract}

\section{Introduction}
Low energy supersymmetry (SUSY)\cite{Wess:1992cp} has been playing a
central role in the quest for physics beyond the
standard model (SM).  Since phenomenological consistency requires SUSY
to be broken, and broken softly (so as not to reintroduce any
quadratic divergence), the Lagrangian of the Minimal Supersymmetric
Standard Model (MSSM)\cite{HaberKaneEtc,KaneKingRev} includes soft and
gauge invariant SUSY breaking terms.  While the generic MSSM
Lagrangian may contain many arbitrary soft terms, specific models for
SUSY breaking have been proposed that provide relationships 
between the MSSM parameters.  Incorporating well-motivated new
interactions and particles at high mass scales, such scenarios 
drastically reduce the large number of unknown parameters in the MSSM
to only a few, thereby making the model more predictive.  We will
focus here only on supergravity
(SUGRA)\cite{Chamseddine:1982jx,appliedPN} type of models where SUSY
is considered as a local symmetry.  These models incorporate a hidden
sector wherein SUSY is broken, and a visible sector where the MSSM
fields reside and to which the breaking is communicated by
gravitational interactions.  In $N=1$ SUGRA, which incorporates grand
unification, one has a choice of three functions in building a
model\cite{Chamseddine:1982jx,appliedPN,coupsug}, namely the gauge
kinetic energy function $f_{\alpha \beta}(z_i)$, the
K\"ahler potential $K(z_i,z_i^\dagger)$, and the superpotential
$W(z_i)$, where $z_i$ refer to matter fields.  In mSUGRA, the minimal
version of the model, one has a flat K\"ahler potential and a flat
gauge kinetic energy function.  The corresponding soft SUSY breaking
sector is characterized by only a few parameters, normally specified
at the scale of the grand unified theory (GUT) 
viz. $M_G \sim 2 \times
10^{16}$~GeV\cite{Nath:1983aw,Hall:1983iz}.  These are the universal
gaugino mass $\mhalf$, the universal scalar mass $m_0$, the universal
trilinear coupling $A_0$ and the universal bilinear coupling $B_0$.
In addition to these, there is a superpotential parameter, namely the
Higgs mixing term $\mu_0$. Unlike in the SM, where the breaking of the
electroweak symmetry necessitates the explicit introduction of a
negative valued scalar mass-squared, in a generic SUGRA model, the
said breaking can be realized even for a positive mass-squared term in
the bare Lagrangian, thanks to radiative
corrections\cite{Chamseddine:1982jx}.  In other words, the
renormalization of the soft SUSY breaking terms as one moves from the
unification scale down to the electroweak scale automatically engenders a
negative mass-squared thereby breaking the
symmetry\cite{grandunisug,inoue,oldrge,newrge}.  In a similar vein,
the low energy parameters of the MSSM (which are quite large in
number) are obtained from only a few unification scale parameters via
the renormalization group equations (RGE)\cite{newrge} integrated from
$M_G$ to the electroweak scale ($\sim M_Z$).  The two minimization
conditions for the Higgs potential then eliminate $\mu_0$ (except for
its sign) on the one hand, and, on the other, relate $B_0$ to
$\tan\beta$~($\equiv \langle H_U \rangle / \langle H_D\rangle$), the
ratio of Higgs vacuum expectation values.  Thus mSUGRA may be
characterized by $\tan\beta$, $m_{1/2}$, $m_0$, $A_0$ and 
sign($\mu$)\footnote{Our choice of sign for $\mu$ and 
$A_0$ follows the standard 
convention of Ref\cite{signconvention}.}.
With all the low energy parameters of the MSSM being generated in
terms of these few parameters, one has a considerable amount of
predictivity for the MSSM spectrum.

A different problem remains though, namely that of the SUSY CP
violating phases.  Many phases of SUGRA models can be rotated away. In
an universal scenario like mSUGRA, the gaugino masses can be
considered real with the result that only two combinations of phases
(beyond the Cabibbo-Kobayashi-Maskawa quark mixing (CKM) 
phase already present in the SM) are physical. A
convenient choice for the two is given by $\phi_{A_0}$ for $A_0$ (at
$M_G$) and $\theta_B$ for the $B$-parameter at the electroweak scale.  
It should be noted though that many analyses prefer to work with $\theta_\mu$, the
phase of $\mu$, instead of $\theta_B$.  An advantage of this latter
choice is that $\theta_{\mu_0} \sim \theta_\mu$ since $\theta_\mu$
does not run up to the one-loop level.  These different descriptions
can be understood in terms of $U(1)_R$ and $U(1)_{PQ}$~(Peccei-Quinn)
symmetries and the choice of reparametrization invariant combinations
of phases, a discussion of which may be found in
Refs.\cite{KaneKingRev,Ibrahim:1998je}. A selection of past analyses
using $\theta_B$ as an input parameter may be seen in
Refs.\cite{Accomando:1999uj,Arnowitt:2001pm,Accomando:1999zf,
Barger:2001nu,Falk:1998pu,Garisto:1996dj}.  Here we note that a choice
of $\theta_B$ instead of $\theta_\mu$ as a phase parameter makes the
entire set of input parameters to be of soft-breaking origin.

A few important points need to be noted in the context of the SUSY CP
problem.  The latter arises from the fact that the phases are highly
constrained by the experimental limits on the electric dipole moments
(EDM) of the electron and the
neutron\cite{Ibrahim:1998je,Accomando:1999uj,Arnowitt:2001pm,Accomando:1999zf,
Barger:2001nu,Falk:1998pu,EDMmany1,EDMmany2}. Consequently, we are forced 
to admit one of the three eventualities:

\begin{enumerate}
\item The phase $\theta_B$ is very small---$O(10^{-2})$ or
$O(10^{-3})$---if the superpartners are not considered to be very
heavy\footnote{$\theta_B$ may reach up to $\sim 0.1$ in the focus 
point zone\cite{focuspointEDM}.}.  In addition, the phases of the $A$-parameters at the
electroweak scale are also constrained.  In mSUGRA with phases, the
requirement of having a very small $\theta_B$ typically translates
into a relatively large but {\em highly fine-tuned} value for ${\rm
arg}(B_0)$ (i.e., $B$ at $M_G$).  This, in turn, constrains the phase
$\phi_{A_0}$ of $A_0$, although to a somewhat 
lesser degree.  The fact that the issue of fine-tuning in phases at
the GUT scale arises out of the combined requirement of satisfying the
EDM constraints and the radiative electroweak symmetry breaking was
discussed in great detail in
Refs.\cite{Accomando:1999uj,Accomando:1999zf,Arnowitt:2001pm} as well
as in Refs.\cite{Barger:2001nu,Falk:1998pu}.  In this paper we try to
focus our attention on this problem by looking at suitable models
beyond mSUGRA that can have unique features in the evolution of $B$.

\item The phases are large and less fine-tuned but the sparticles are
massive. Of course, fully ameliorating the SUSY CP problem in this
fashion requires that the sfermions be super-massive, thereby
aggravating the problem of the little mass hierarchy in the Higgs
sector.  We will investigate whether the amount of fine-tuning can be
reduced even while one considers a lighter sparticle spectra.

\item Finally there is the possibility that the SUSY breaking
parameters may have special pockets where there can be a large amount
of internal cancellations between the diagrams contributing to the
electric dipole moments of electron and neutron\cite{EDMmany1}.  This
means that phases could be large while sparticle masses are
significantly light.  This scenario is highly parameter dependent
and clearly depends on very delicate cancellations.  
Hence we will not include this in our work while trying to focus on
generic behaviors.
\end{enumerate}

As mentioned above, we would like to address the first and the second
issues in this analysis. We are particularly interested in exploring
the possible role of non-universal gaugino masses (NUGM) in reducing
the fine-tuning in the phase $\thetab0$.  To quantify the latter, we
consider a {\it naturalness like} measure of the form 
\beq
\Phi=[{\Delta \thetab0}/{\Delta\theta_B}]_{\theta_B \rightarrow 0}.
\label{sensitivityPhi}
\eeq 
A large value for $\Phi$ would mean a lesser degree of fine-tuning of
$\thetab0$ with respect to a variation in $\theta_B$ satisfying the
EDM constraints.  The phase-derivative is evaluated at $\theta_B \sim
0$ with the choice being dictated by the fact that the EDM constraints
force $|\theta_B|$ to be close to zero.  Thus, this is a restrictive
definition compared to the type of fine-tuning defined in
Ref.\cite{Arnowitt:2001pm}.
    
We will see that the issue of such fine-tuning of phase can be
addressed by focusing on scenarios where there is a large evolution
of the bi-linear Higgs coupling parameter $B$ between the electroweak
scale and the GUT scale.  The evolution of $B$ depends on the $U(1)$
and the $SU(2)$ gaugino masses, the trilinear couplings and
$\tan\beta$.  Within mSUGRA, in addition to the evolution of $|B|$
being typically small, the phase $\thetab0$ also turns out to be quite
fine-tuned (i.e. $\Phi$ tends to be small).  In other words, for a
given $\thetab0$ satisfying the EDM constraints, the variation $\Delta
\thetab0$ that still is consistent with the constraints is generally
much smaller than the variation $\Delta \theta_B$ allowed at the
electroweak scale\cite{Accomando:1999uj}.  As we will see, the
evolution in $|B|$ may be enhanced by appropriate mass relationships
between the gauginos that are away from universality at $M_G$.  At the
same time, these would help in reducing the above-mentioned fine-tuning
so that $\Phi$ can be significantly increased in specific NUGM
scenarios.

We, however, desist from choosing
an arbitrary non-universal gaugino
mass scenario since that will introduce new
phases\cite{Accomando:1999zf}.  As we will see in Sec.\ref{NUGMdesc},
non-universalities in gaugino masses may originate from a non-trivial
gauge kinetic energy function. The latter is a function of chiral
superfields and transforms as a symmetric product of the adjoint
representations of the underlying gauge group. This leaves $f_{\alpha
\beta}$ with the possibility of being in one or more of several
representations, one of which is the singlet.  While the choice of the
singlet corresponds to mSUGRA, the non-singlet representations give
rise to non-universalities in the gaugino masses.  It is possible to
identify a suitable non-singlet representation in isolation ({\em
i.e.}, we will not combine a non-singlet representation with the
singlet or other non-singlet representations) whose gaugino mass
pattern is effective in generating a large evolution in $B$.  At the
same time, there will be no additional phases to worry about since the 
overall phase
of the gaugino masses can be rotated away in a fashion similar
to that in mSUGRA.

In this paper, we will analyze the consequences of a large evolution
of the $B$-parameter (mostly in the presence of such
non-universalities) on the CP violating phases.  Here, the basic input
parameters are $\tan\beta$, $m_0$, $\mhalf$ (providing with definite
NUGM patterns), $|A_0|$ along with its phase $\phi_{A_0}$ and the
phase $\theta_B$ of $B$ given at the electroweak scale ($\sim M_Z$).
Note that $|B|$ at the electroweak scale is obtained via radiative
electroweak symmetry breaking (REWSB) condition. Subsequently, $|B_0|$,
the GUT scale magnitude of the $B$-parameter 
along with its phase $\thetab0$ 
is obtained via RGEs.  We will identify broad but correlated regions
of parameter space where there can be a significant degree of
reduction of the phase sensitivity while going from mSUGRA to a type
of NUGM models.
   
    The paper is organized as follows. In Sec.\ref{NUGMdesc}, we 
discuss the non-universal gaugino mass models.  The study of the
relevant contributions from different sectors in the associated RGEs
of $B$ and $A$ parameters allows us to identify the non-singlet
representations which provide with a large evolution in $B$.  We will
probe the parameter space that is suitable for reducing the amount of
fine-tuning in the CP violating phases.  In Sec.\ref{results},  we present
the numerical results for the evolution of $B$.  
An analysis in the absence of phases points us to the 
favored regions of parameter spaces. On inclusion of phases, 
this facilitates the identification of the regions with significantly 
reduced level of fine-tuning, Finally, we conclude in Sec.\ref{secConclusion}.

\section{Non-universal gaugino masses and enhanced evolution of $B$}
\label{NUGMdesc}
Non-universality in gaugino masses may originate from a non-trivial
gauge kinetic energy function $f_{\alpha \beta}$ which, in turn, is a
function of the chiral superfields in the theory.  The indices
$\alpha, \beta$ run over the generators of the gauge group (for
example, $\alpha=1,2,\dots 24$ for SU(5)).  The gaugino mass matrix is
given by
\begin{equation}
M_{\alpha\beta}= \frac{1}{4}\bar{e}^{G/2}G^a(G^{-1})^b_a
(\partial f^*_{\alpha\gamma}
/\partial z^{*b})f^{-1}_{\gamma\beta}
\end{equation}
where $G = -\ln[\kappa^6 W W^*]- \kappa^2 K$. Here, $W$ is the
superpotential, $K(z,z^*)$ is the K\"ahler potential, $z^a$ are the
complex scalar fields, and $\kappa = (8\pi G_N)^{-\frac{1}{2}} =
0.41\times 10^{-18}$ GeV$^{-1}$ with $G_N$ being Newton's constant.
The functions $f_{\alpha \beta}$ may have non-trivial field
contents, or in other words, may contain combinations of field
transforming as either singlet or non-singlet irreducible
representations\cite{eent}.  With the gauginos being Majorana
particles, $f_{\alpha \beta}$, of necessity, must be contained in the
symmetric product of the adjoint representations of the gauge group.
For example, in the case of SU(5),
\begin{equation}
f_{\alpha \beta} \supset (24 \otimes 24)_{sym}=1 \oplus24 \oplus 75 \oplus 200   \ .
\label{su5breakup}
\end{equation}  
For the singlet case, one has $f_{\alpha \beta}=\delta_{\alpha \beta}$
which indeed leads to universality of gaugino masses. Similarly, the
non-singlet representations will give rise to non-universal gaugino
masses. 
 
In general $M_i(M_G)=\mhalf \sum_r C_r n_i^r$, where $C_r$'s give the
relative weights of each contributing representation 
and $n_i^r$, for the subgroup $i$, are essentially the 
Clebsch-Gordan coefficients 
corresponding to the breaking by the adjoint Higgs 
field\cite{eent,corsetti2,hill}.
For the case of $SU(5)$,
the coefficients $n_i^r$ are displayed in Table~\ref{tabrelativeweights}. 
Clearly, the non-singlet representations have characteristic mass relationships
for the gaugino masses at the GUT scale.  Past analyses exploring
various phenomenological implications of such non-universality may be
found in Refs.\cite{eent,corsetti2,andersonNonuniv,ucnonuniv,nonunivso10}.

As we shall argue later, the adjoint representation $r=24$ for $f_{\alpha
\beta}$ (NUGM:24 in the notation of Table~\ref{tabrelativeweights}) is
the most interesting one in the context of the present investigation.
Consequently, we will analyze this case in isolation, or, in
other words, assume that the sole contribution to $f_{\alpha \beta}$
is from a 24-plet structure. Apart from reducing the number of free
parameters, this has the additional advantage that no new phase
degree of freedom for the gaugino masses is introduced.  With the 
gaugino mass ratios at the GUT scale now being given by
$M_3(M_G):M_2(M_G):M_1(M_G)=1:-3/2:-1/2$, for a positive
gluino mass, the other two gaugino mass parameters are negative, 
a signature different from mSUGRA. This
indeed would turn out to be useful in our quest.  As mentioned
earlier, we only consider either $C_1 = 1$ (mSUGRA) or $C_{24}=1$
(NUGM:24) with all other $C_r$'s assumed to be zero.

\begin{table}[!h]
%\vspace{3mm}
{\centering
\begin{tabular}[hbt]{|c|c| c c c|}
\hline
$r$ & Label & $M^G_3$ & $M^G_2$ & $M^G_1$ \\
\hline
\hline
1 &  mSUGRA & 1 & 1 & 1 \\
24 & NUGM:24 & 2 & $-3$ & $-1$\\
75 & NUGM:75 & 1 & 3 & $-5$  \\
200 & NUGM:200 & 1 & 2 & 10  \\
\hline
\end{tabular}
\par}
\centering
\caption{\em The coefficients $n_i^r$ as pertaining to the 
 $SU(3)$, $SU(2)$ and $U(1)$
gaugino masses at the GUT scale for different representations 
of $SU(5)$.}
\label{tabrelativeweights}
\end{table}

An analogous analysis with SO(10) as the underlying gauge group is
also possible\cite{chamoun,nonunivso10}, though we will not
investigate it in this paper.  Similar to Eq.\ref{su5breakup} here,
one has $(45 \times 45)_{sym}=1+54+210+770 $.  If the symmetry
breaking pattern is $SO(10) \rightarrow SU(4) \times SU(2) \times
SU(2) \rightarrow SU(3) \times SU(2) \times U(1)$, one finds from the
$54$-plet that $M_3(M_G):M_2(M_G):M_1(M_G)=1:-3/2:-1$.  
This pattern 
is quite similar to NUGM:24 as can be ascertained from
Table~\ref{tabrelativeweights}.  We would like to comment at this
point that, in general, such non-universal gaugino mass scenarios change
the gauge coupling unification conditions\cite{eent,hill}.  However, it is
still possible to find specific conditions\cite{ucpnbtau,eent} under
which the usual gauge coupling unification condition remains unaltered
and we consider this in our work. Note though that our results are
quite robust and have very little dependence on the exact details of
the spectrum.

\subsection{Nature of evolution of $B$ with real parameters}
\label{AnalysisSubSec1}
We now identify the differences between mSUGRA and NUGM:24 in regard
to the evolution of the $B$-parameter {\em in the absence of CP
violating SUSY phases}.  This, in turn, will help us in understanding
the evolution of $\theta_B$ upon the inclusion of the phases (see
Refs.\cite{Accomando:1999zf,Barger:2001nu,Falk:1998pu,Garisto:1996dj}
for past analyses discussing phase evolutions).  Note that $\mu^2$ and
$B$ are determined via the REWSB condition, viz.
\begin{equation}
\begin{array}{rcl}
|\mu|^2 & = & \displaystyle
-\frac{1}{2} M^2_Z +\frac {m_{H_D}^2-m_{H_U}^2 \tan^2\beta} {\tan^2\beta -1}
+ \frac {\Sigma_1 -\Sigma_2 \tan^2\beta} {\tan^2\beta -1} 
\\[2ex]
\sin(2\beta)& = & 2|B\mu|/(m_{H_D}^2+m_{H_U}^2+2\mu^2+\Sigma_1+\Sigma_2) \ ,
\label{EqnBewsb}
\end{array}
\end{equation}
where $\Sigma_i$ represent the one-loop
corrections~\cite{effpot1,effpot2}.  The Higgs scalar mass parameters
$m_{H_D}$ and $m_{H_U}$, and thereby $\mu^2$ and $B$ depend 
quite strongly on $m_0$ as well as on $\mhalf$.  To one-loop order, the running of the
$B$ parameter has two additive components, the first proportional to
the gaugino masses and the second depending on a combination of the trilinear
couplings and the Yukawa couplings~\cite{Accomando:1999zf,newrge}, namely, 
\begin{equation}
{\frac{dB}{{dt}}}= (3\tilde{\alpha}_2{\tilde{m}_2}
+\frac 35\tilde { \alpha}_1
{\tilde{m}_1}) +(3Y_tA_t +3Y_bA_b+Y_\tau A_\tau) \ ,
\label{EqnBrge}
\end{equation} 
where $t=ln(M_G^2/Q^2)$ with $Q$ being the renormalization scale.  $\tilde
\alpha_i=\alpha_i/(4\pi)$ are the scaled gauge coupling
constants (with $\alpha_1=\frac53 \alpha_Y$) and $\tilde m_i$ for
$i=1,2,3$ are the running gaugino masses. Furthermore, $Y_i$ represent
the squared Yukawa couplings, e.g, $Y_t \equiv y_t^2/(4\pi)^2$ where
$y_t$ is the top Yukawa coupling.  In a similar vein, the
evolution of the trilinear terms is given by
\begin{equation}
\begin{array}{rcl}
 \displaystyle
{\frac{dA_t}{{dt}}} & =& \displaystyle
-(\frac {16}{3}\tilde{\alpha}_3{\tilde{m}_3}
+3\tilde{\alpha}_2{\tilde{m}_2}
+\frac {13}{15}\tilde { \alpha}_1
{\tilde{m}_1}) -6Y_tA_t -Y_b A_b  \\[1.5ex]
 \displaystyle
{\frac{dA_b}{{dt}}} & =& 
 \displaystyle -(\frac {16}{3}\tilde{\alpha}_3{\tilde{m}_3}
+3\tilde{\alpha}_2{\tilde{m}_2}
+\frac {7}{15}\tilde { \alpha}_1
{\tilde{m}_1}) -Y_tA_t -6Y_b A_b-Y_\tau A_\tau  \\[1.5ex]
 \displaystyle
{\frac{dA_\tau}{{dt}}} & =& 
 \displaystyle -(3\tilde{\alpha}_2{\tilde{m}_2}
+\frac {9}{5}\tilde { \alpha}_1{\tilde{m}_1})
-3Y_b A_b-4Y_\tau A_\tau \ .
\label{EqnAtrge}
\end{array}
\end{equation}

For small $\tan\beta$, the contributions from the bottom quark and tau Yukawa
couplings $y_b$ and $y_\tau$ may be neglected, and the RGEs 
approximately integrated to obtain~\cite{Accomando:1999uj}
\begin{equation}
B - B_0 \simeq \frac{D_0(t) - 1}{2} \,  A_0 - C(t) \, \mhalf  \ ,
\label{EqnBzVsB0}
\end{equation}
where $D_0(t) \equiv 1-6Y(t)F(t)/E(t)$ with $t$ 
corresponding to the electroweak scale.  The functions $E(t)$ and $F(t)$ 
encapsulate the running of the gauge coupling constants, viz,
\[
\begin{array}{rcl}
E(t) &= & \displaystyle
(1+\beta_3t)^{16/(3b_3)}(1+\beta_2t)^{3/b_2}(1+\beta_1t)^{13/(15b_1)}
\\[2ex]
F(t) & \equiv & \displaystyle \int_{0}^{t} E(t') dt'
\end{array}
\]
where $\beta_i=b_i {\tilde \alpha}_i(0)$ and
$(b_1,b_2,b_3)=(33/5,1,-3)$ are the coefficients in the respective
one-loop beta-functions. Of course, unification imposes the boundary
condition that $\alpha_i(0)=\alpha_G \sim 1/24$. At the top mass scale 
($Q=m_t$), $D_0\simeq 1-(m_t/200\sin\beta)^2 \lsim 0.2$ is indeed 
a very good approximation. The function $C(t)$, in
Eq.\ref{EqnBzVsB0}, on the other hand, is given by
\begin{equation}
C(t)=-\frac{1}{2}(1-D_0){H_3 \over F} +\left( 3h_2+ \frac{3}{5} h_1 \right)
{\alpha_G \over {4\pi}} \ ,
\label{EqnForC}
\end{equation}
where
\[
\begin{array}{rcl}
h_i(t)& \equiv & \displaystyle\frac{t} {(1+\beta_i t)}
\\[2ex]
H_3(t)& \equiv & \displaystyle \int_{0}^{t} {E(t') H_2(t')}dt'
\\[2ex]
H_2(t) & \equiv &  \displaystyle {\tilde\alpha}(0) 
 \left({\frac{16}{3}} h_3 +3h_2 +
{\frac{13} {15}} h_1  \right)  \ .
\end{array}
\]
For the generic (NUGM) case, the above results remain the same except
that~\cite{corsetti2} 
\beq h_i(t) \longrightarrow {\tilde h}_i(t)
\equiv h_i(t) \, {\frac {{\tilde m}_i(0)} {\mhalf}}  \ .
\eeq

Note that, in $d B / d t$, the gaugino contribution is positive for
mSUGRA, but negative for NUGM:24. Thus, it is useful to understand the
nature of evolution of trilinear couplings in either scenario so as to
evaluate their role in the evolution of $B$.  For the mSUGRA case, the
gaugino contributions to $d A_i / d t $ are always negative (vide
Eq.\ref{EqnAtrge}). Hence, it is obvious that if $A_0$ not be too
large, then $A_i$ would typically turn negative by the electroweak
scale. In fact, the large gluino contributions render both $A_t$ and
$A_b$ negative well above the electroweak scale. This implies, that in
this case (mSUGRA), the two pieces in $d B / d t$ would tend to cancel
each other, an effect also manifested by the  
smallness of $C$ in Eq.\ref{EqnBzVsB0}. In turn, this leads to a 
small value for $\Delta B \equiv |B_0-B|$ in mSUGRA.

Comparing the evolution of the trilinear terms in NUGM:24 with 
that in mSUGRA, it turns out that a qualitative difference arises
only in the case of $A_\tau$, while for $A_t$ and $A_b$ the difference 
between the scenarios is only a quantitative one. This is easy to understand
given the overwhelming dominance, in the last two cases, of the gluino 
contribution over those from the electroweak gauginos. 
Specifically, for $A_0=0$,
$A_\tau$ at the weak scale comes to be negative for mSUGRA while it is
positive (with usually a larger magnitude) for NUGM:24.  
Given the relative weights of the $A_i$ terms in Eq.\ref{EqnBrge}, 
it is thus quite apparent that the total contribution from the trilinear 
couplings to the evolution of $B$  is quite
similar in the two models.  On the other hand, since the signs of
${\tilde{m}_{1,2}}$ are reversed in NUGM:24, the aforementioned
cancellations in $d B / d t$ would no longer be operative; rather, the
different contributions would enhance each other leading to a large
$\Delta B$. This is the very reason why we choose to concentrate on
models like NUGM:24.  We note in passing that although the RGE for $B$
does not explicitly include the SU(3) gaugino mass, it implicitly
depends on the latter via the contributions from trilinear couplings.

We now discuss the dependence of $B$ and $B_0$ on $m_0$ and the 
other parameters.  Being obtained from the REWSB condition of
Eq.\ref{EqnBewsb}, $B$ (and hence $B_0$) evidently depends on $m_0$
quite strongly. The structure of Eq.\ref{EqnBrge} suggests that, to
one-loop order, $\Delta B$ should not depend on $m_0$.  However, a
subsidiary dependence arises through the determination of the scale at
which the minimizations of Higgs potential (i.e. REWSB) is to be
performed.  Canonically, this scale is determined by demanding that
the contribution, to $\mu^2$, of the 1-loop correction terms of the
effective potential be small.  In our analysis this scale is
approximately halfway between the lowest and highest mass of the
spectra and, generally, is not very far from the average stop mass
scale $\sqrt{m_{{\tilde t}_1}m_{{\tilde t}_2}}$ (see
Ref.\cite{Chan:1997bi}).  Since this scale does depend on $m_0$, it
leads to a small dependence in $\Delta B$ as well by virtue of being a
limit of integration for the RGEs.

\subsection{Incorporating CP violating phases: $|B_0|/|B|$ and phase 
naturalness measure $\Phi$} 
\label{AnalysisSubSec2}
Even on inclusion of phases for the $A$ and $B$
parameters, the RGEs formally remain the same as in
Eqs.(\ref{EqnBrge}\&\ref{EqnAtrge}).  The evolution of the phases can
then be extracted by comparing the real and imaginary parts of the said
equations.  Clearly, unlike in the case of the real parts, the
imaginary parts of the beta functions for $A$'s and $B$ do not depend
on the gaugino masses and hence there is no cancellation between the
different contributions.  Furthermore, even a vanishing $\thetab0$ can
lead to a non-zero $\theta_B$ provided $A_0$ has a non-trivial
phase. For example, in the small $\tan\beta$ limit, the explicit
analytical solution gives 
\begin{equation} \begin{array}{rcl}
|B|\sin\theta_B & = & \displaystyle
|B_0|\sin\thetab0-\frac{1}{2}(1-D_0)|A_0|\sin\phi_{A_0} \\[2ex]
|B|\cos\theta_B & = & \displaystyle
 |B_0|\cos\theta_{B_0} - \frac12(1 - D_0)|A_0|\cos\phi_{A_0} 
-C\mhalf \ .
\label{EqnBimag}
\end{array}
\end{equation}

We examine now the interdependence between the phases, their evolution
(also see Ref.\cite{Accomando:1999uj}) and the phase sensitivity
$\Phi$ for different values of $\tan\beta$ and other parameters both
within mSUGRA as well as NUGM:24.  As we have already mentioned, the
EDM constraints limit $\theta_B$ to be tiny ($\lsim 0.1$, and
typically much smaller). Now, if 
either of $|A_0|$ or $\phi_{A_0}$ is small
(actually, if $|A_0|\sin\phi_{A_0} \ll |B|\sin\theta_B$), then $\thetab0$
would be determined essentially by $|B|$, $|B_0|$ and $\theta_B$.
In this case, 
$\phi_{A_0}$ would be quite unconstrained.  The
dependence on $\tan\beta$ is crucial and is best understood by
considering the two opposite limits, namely small and large values:

\begin{itemize} 
\item For a small $\tan\beta$ ($\lsim 5$ or so),
$\sin2\beta$ is large, and therefore $|B|$ is appreciably large (see
Eq.\ref{EqnBewsb}).  Within mSUGRA, for not too large a value of
$|A_0|$, the GUT scale value $|B_0|$ is then quite comparable to $|B|$.
This can be understood by recognizing the cancellations
between the various terms in Eq.\ref{EqnForC} that keeps $C$ small and
thereby keep $B - B_0$ relatively small (courtesy
Eq.\ref{EqnBzVsB0}).  Consequently, in such a scenario, 
$\thetab0$ is not too different from $\theta_B$.  
This remains true even for
$\phi_{A_0}=\pi/2$ which maximizes the EDM values\cite{EDMmany2}.  

On the contrary,
the situation in NUGM:24 is quite different. Here, a larger difference
between $|B|$ and $|B_0|$ is generated by the enhancement in $C$.
Consequently, $\thetab0$ becomes appreciably different 
from (and numerically larger than) $\theta_B$.

\item For a large value of $\tan\beta$, on the other hand, $\sin\,2\beta$
is quite small.  Thus, unless $|\mu|$ is extremely tiny (as happens,
for example, in hyperbolic branch/focus point\cite{Chan:1997bi,HBFP}
scenarios), $|B|$ is constrained to be small and has only sub-dominant
influence on the evolution of $\theta_B$.  This, in turn, implies that 
the value of
$\thetab0$ becomes strongly correlated with that of $\phi_{A_0}$. In
other words, a high degree of fine-tuning in one will necessitate a 
similar degree of fine-tuning in the other.  
\end{itemize}

We now focus on the issue of phase sensitivity. As Eq.\ref{EqnBimag}
suggests, the range allowed to $\theta_B$ (i.e. $\Delta \theta_B$)
imposes rather strong limits in the $\thetab0$--$\phi_{A_0}$
plane. Adopting the measure of phase naturalness $\Phi$ (as espoused in
Eq.\ref{sensitivityPhi}), one may estimate, from Eq.\ref{EqnBimag}, 
the amount of fine-tuning
associated with the phase $\thetab0$. Now, as the RGEs suggest, the
implicit dependence of $\Phi$ on $A_0$ occurs primarily through the
dependence of $B_0$ itself on $A_0$. Thus, to the leading
order, one has an approximate relation of the form\cite{Accomando:1999uj}
\begin{equation}
\Phi \sim |B \, / \, B_0| \ .
\label{approxPhi}
\end{equation}
We would like to point out that although the above simplification (as
also those of neglecting $y_b$ and $y_\tau$) is quite illustrative, we
do not take recourse to it.  Rather we solve the complete set 
of RGEs numerically 
and also compute $\Phi$ numerically directly from its definition 
(Eq.\ref{sensitivityPhi}).

Note that, as obtained from Eq.\ref{sensitivityPhi} and the first of
Eqs.\ref{EqnBimag}, the measure $\Phi$ actually involves a factor of
$\cos\theta_{B_0}$ in the denominator. This causes $\Phi$ to be very
large when $\theta_{B_0}$ is close to $\pi/2$, as also a change of
sign for $\Phi$ when $\theta_{B_0}$ crosses $\pi/2$. We will see that
this is indeed the case for NUGM:24 where $\theta_{B_0}$ can easily cross 
$\pi/2$ owing to a large degree of phase
evolution.  In the mSUGRA scenario, on the other hand, such a feature  
rarely appears.

As we have already discussed, mSUGRA is associated with 
a relatively small degree of evolution in $B$, and hence 
$|B| \sim |B_0|$. This leads to a low value of $\Phi\sim 1$ or, 
equivalently, to a high degree of fine-tuning in $\theta_{B_0}$ . 
On the other hand, a non-universal gaugino mass scenario like NUGM:24
can provide us with a large evolution of $|B|$. This, of course, can
generate either $|B\, / \, B_0| \ll 1$ or $|B \, / \, B_0| \gg 1$. 
The parameter
space corresponding to the latter case (which is typically satisfied
better for smaller $\tan\beta$ zones) reduces fine-tuning in
$\thetab0$.  We will see that the said reduction can be as large as a
factor of 10 to 20 compared to mSUGRA.  And finally, the very same
large evolution of $|B|$ also implies that $|B_0| \sim 0$ could be a 
possibility within such scenarios.  In NUGM:24 where the evolution 
of $B$ is large, the above reduction of $|B_0|$ toward zero is possible when 
$|B|$ is large i.e. when $\tan\beta$ is small.  
In mSUGRA too this is possible, but only to a limited 
degree, as the aforesaid evolution is smaller in extent. 
So $|B|$ needs to be closer to zero in order to have a tiny $|B_0|$.  
In this sense, a requirement of 
a smaller $|B|$ would then favor large values of $\tan\beta$ for mSUGRA.
This we explore numerically in the next section.

\section{Results: Degree of $B$-evolution and phase sensitivity for mSUGRA and NUGM:24}
\label{results}
We show our numerical results in two stages. 
To begin with, we examine the difference between the evolution of $B$
in mSUGRA and the NUGM:24 scenarios in the absence of any phases.
Building on the lessons drawn from this exercise, we investigate next
the core issue at hand, namely the behavior of the phase naturalness
measure $\Phi$ in each of the scenarios and the differences therein.

\subsection{Results in the absence of CP violating phases} 
\label{ResultsSubSec1}
Focusing first on mSUGRA, we begin with the value of $B$ as
determined, by the REWSB conditions, in terms of the other parameters
of the model, viz, $m_0$, $\mhalf$, $A_0$ and $\tan \beta$. 
This study, coupled with that for 
the derived value at the GUT scale, $B_0$,
would serve to indicate the regions of the parameter space for which
the phase sensitivity can be significantly reduced.

\begin{figure}[!h]
\vspace*{-0.05in}                                 
\mygraph{sugBgraphsA}{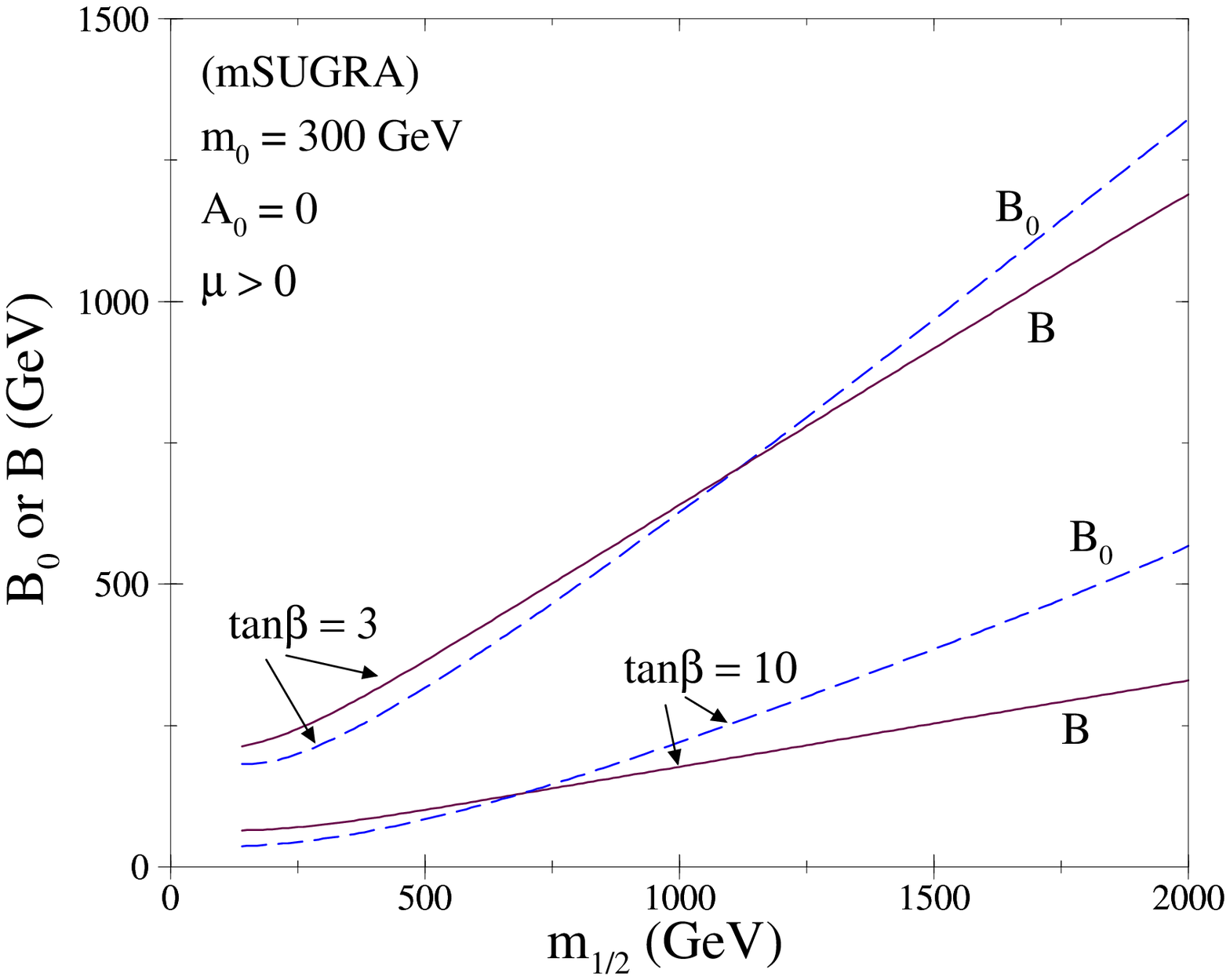}
\hspace*{0.5in}
\mygraph{sugBgraphsB}{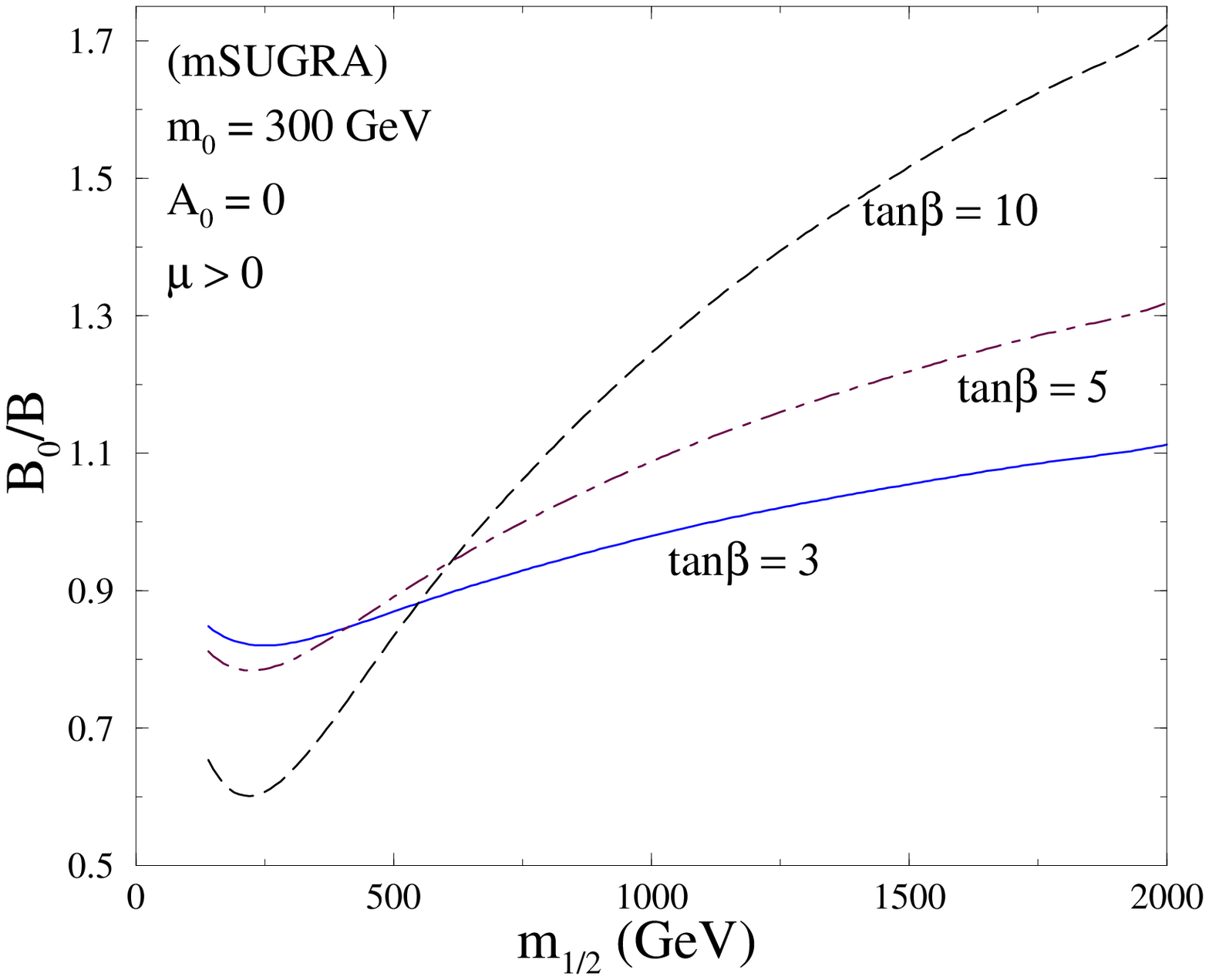}

\mygraph{sugBgraphsC}{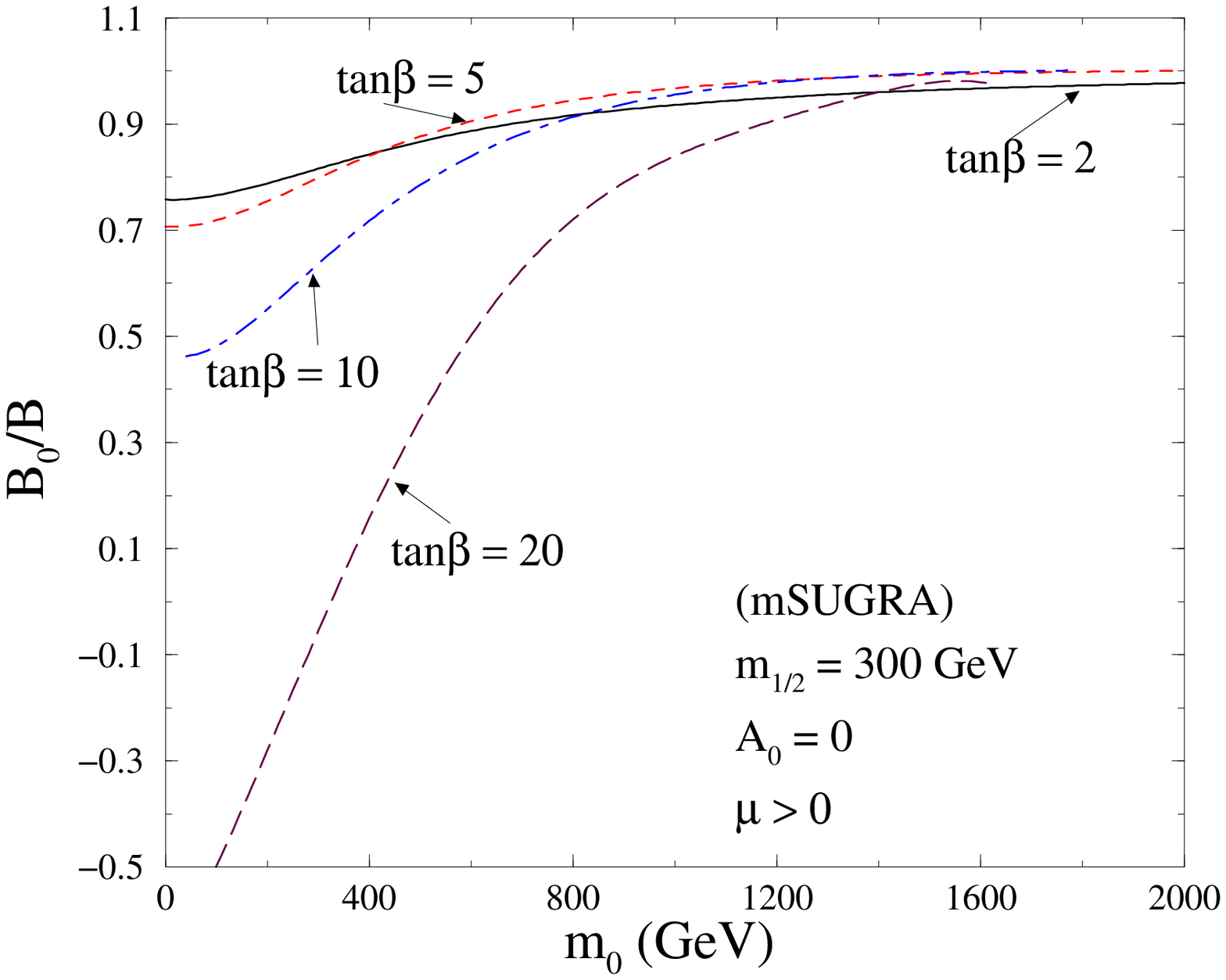}
\hspace*{0.5in}                       
\mygraph{sugBgraphsD}{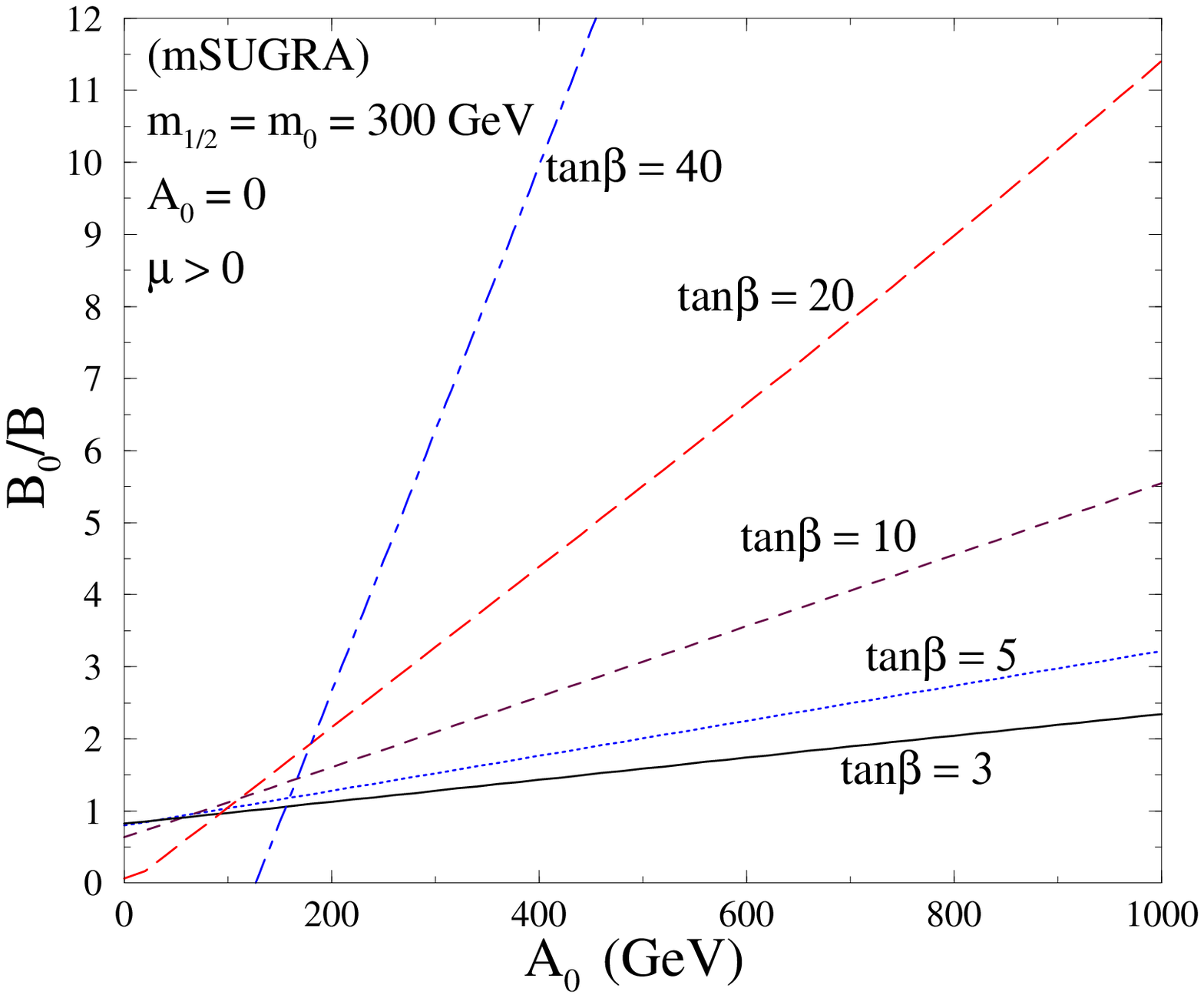}                       
\caption{\em {\em (a)} The dependence of $B$ and $B_0$ on $\mhalf$ in 
mSUGRA with the other parameters fixed as shown. {\em (b,c,d)} The 
dependence of the ratio $B_0/B$ on $\mhalf$, $m_0$ and $A_0$ respectively,
keeping the other parameters fixed.}
\label{sugBgraphs} 
\end{figure} 

Fig.\ref{sugBgraphsA} shows the variation of $B$ and correspondingly
$B_0$ with respect to $\mhalf$. With an illustrative choice of
parameters, viz. $m_0=300$~GeV, $A_0=0$ and $\mu>0$, we exhibit our
results for $\tan\beta=3$ and $10$. One finds that $B$, determined
through the REWSB condition, is almost linear with $\mhalf$.  The
dependence on $\tan \beta$, on the other hand, is quite nonlinear; but
as already touched upon in the previous section, the REWSB condition
implies that, for a given $\mhalf$, $B$ decreases with increase in
$\tanb$.  As for the evolution of $B$, we find that 
$B_0 \sim B$ unless $\mhalf$ is quite large.
This is reflective of the aforementioned cancellations between the
gaugino and trilinear terms of Eq.\ref{EqnBrge}
in mSUGRA.  For our choice of
$A_0=0$, this is same as the cancellations between the terms of $C$ of
Eq.\ref{EqnForC}. Once $\mhalf$ becomes large, the contributions from
the gaugino part of Eq.\ref{EqnBrge} dominates and the cancellations
are no longer as effective.  This causes $B_0$ to supersede $B$ as is
shown in Fig.\ref{sugBgraphsA}.

The information regarding the evolution of $B$ can also be
parametrized in terms of the ratio $B_0 / B$ and this is displayed in
Fig.\ref{sugBgraphsB} as a function of $\mhalf$. This ratio is of
particular interest on account of its relatively straightforward
relation with the phase naturalness measure $\Phi$ 
(note that $\Phi \sim |B \, / \, B_0|$).
As could have been guessed from Fig.\ref{sugBgraphsA} itself, the
variation with $\mhalf$ is nearly monotonic. The shallow dip at small
$\mhalf$ values is a consequence of the variation in the degree of
cancellation between contributions to $d B / d t$ and is difficult to
see analytically from the leading terms alone.  For large $\mhalf$,
the ratio $B_0 / B$ is seen to increase with $\tanb$, while for small
$\mhalf$ the behavior is opposite.  This, within mSUGRA, indicates
that a small value of $\mhalf$, coupled with a large $\tanb$ seems to
be best suited for achieving a low degree of fine-tuning in the
phases.

In Fig.\ref{sugBgraphsC}, we display the dependence of the same ratio
on $m_0$.  While the behavior may seem intriguing at first, note that
$B$ depends on $m_0$ only via the requirement of REWSB.  As
Fig.\ref{sugBgraphsA} has already shown us, for the reference value of
$\mhalf = 300 \rmGeV$, $B_0$ is typically somewhat smaller than
$B$. Now, $B$ grows smaller as $m_0$ decreases. Thus, for small $m_0$
and large $\tan\beta$, $B$ can be very small and the aforesaid
evolution implies that $B_0$ would have been negative.  On the other
hand, for large $m_0$ values, $B$ is large and thus the relatively
small evolution leaves the ratio $B_0 / B$ very close to unity.

The dependence of $B_0 / B$ on the trilinear coupling parameter $A_0$
is quite linear (Fig.\ref{sugBgraphsD}). This, again, can be deduced
from Eq.\ref{EqnBzVsB0} where fixing $\tan\beta$, $\mhalf$ and $m_0$
will give rise to a linear relation between $B_0 / B$ and $A_0$. Note
that progressively larger values for $\tanb$ increases the importance
of the trilinear term contributions to $d B / dt$, thereby increasing
the slope of the curve.

\begin{figure}[!h]
% \vspace*{-1.0in}                                 
\mygraph{nonBgraphsA}{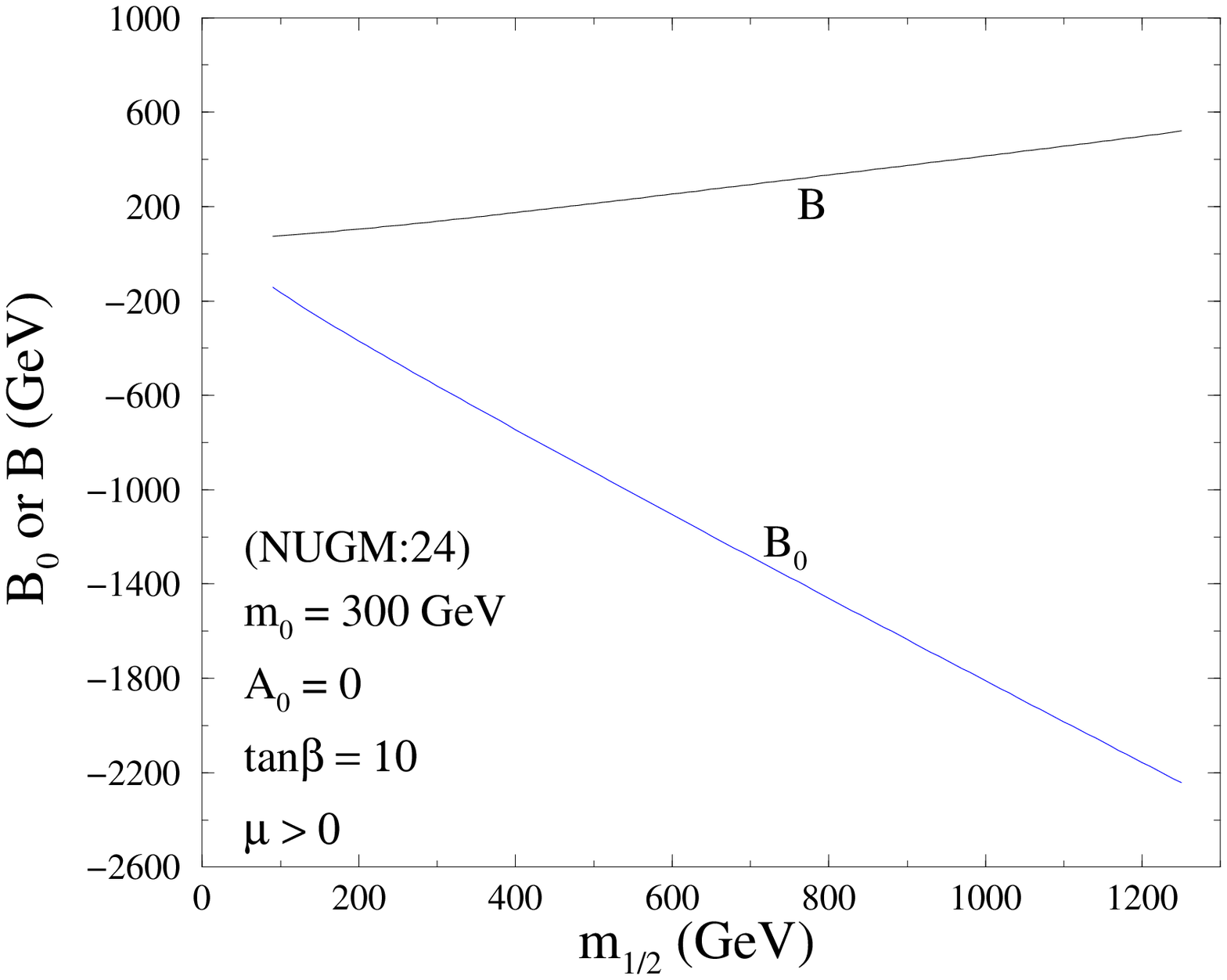}    
\hspace*{0.5in}
\mygraph{nonBgraphsB}{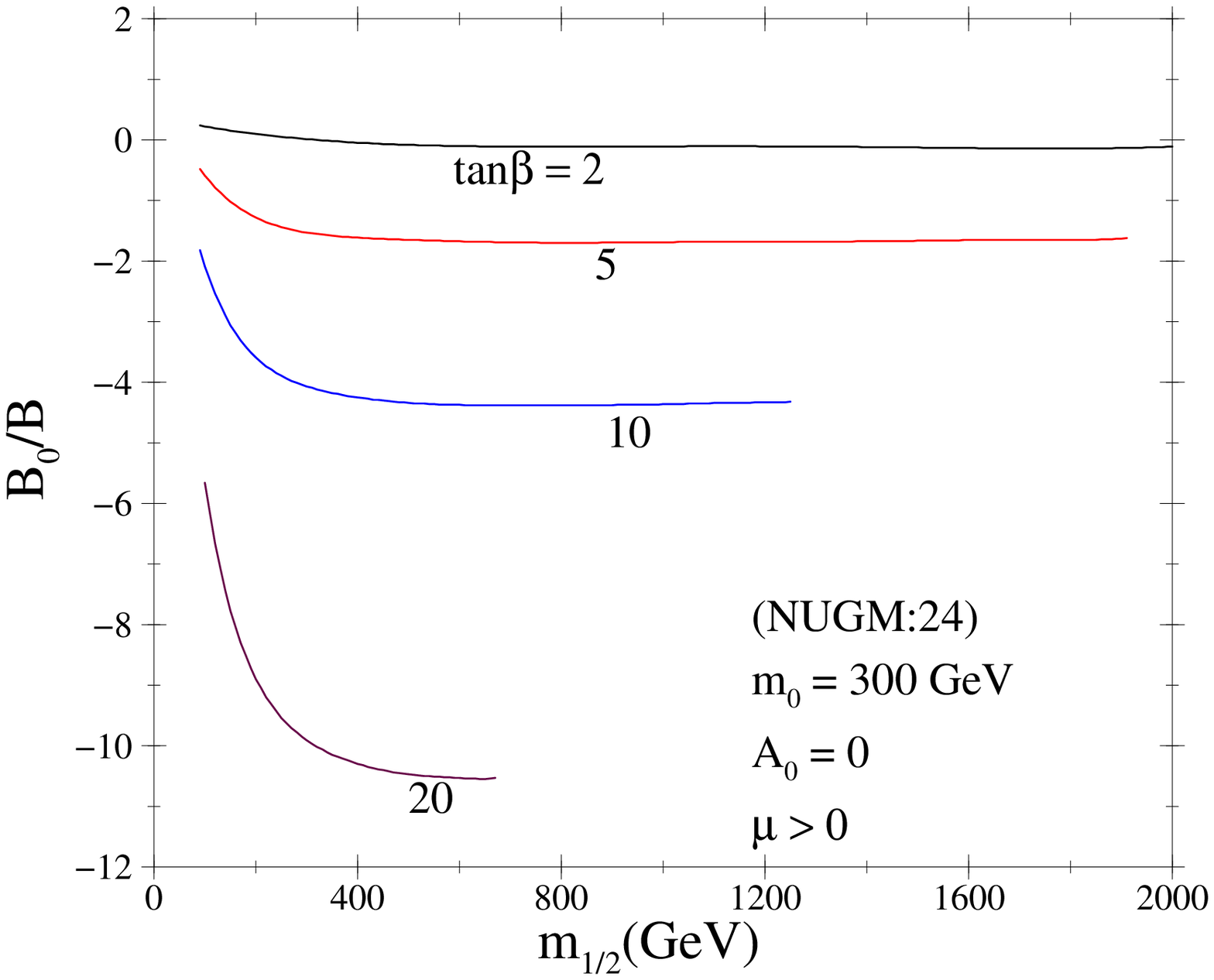} 

\mygraph{nonBgraphsC}{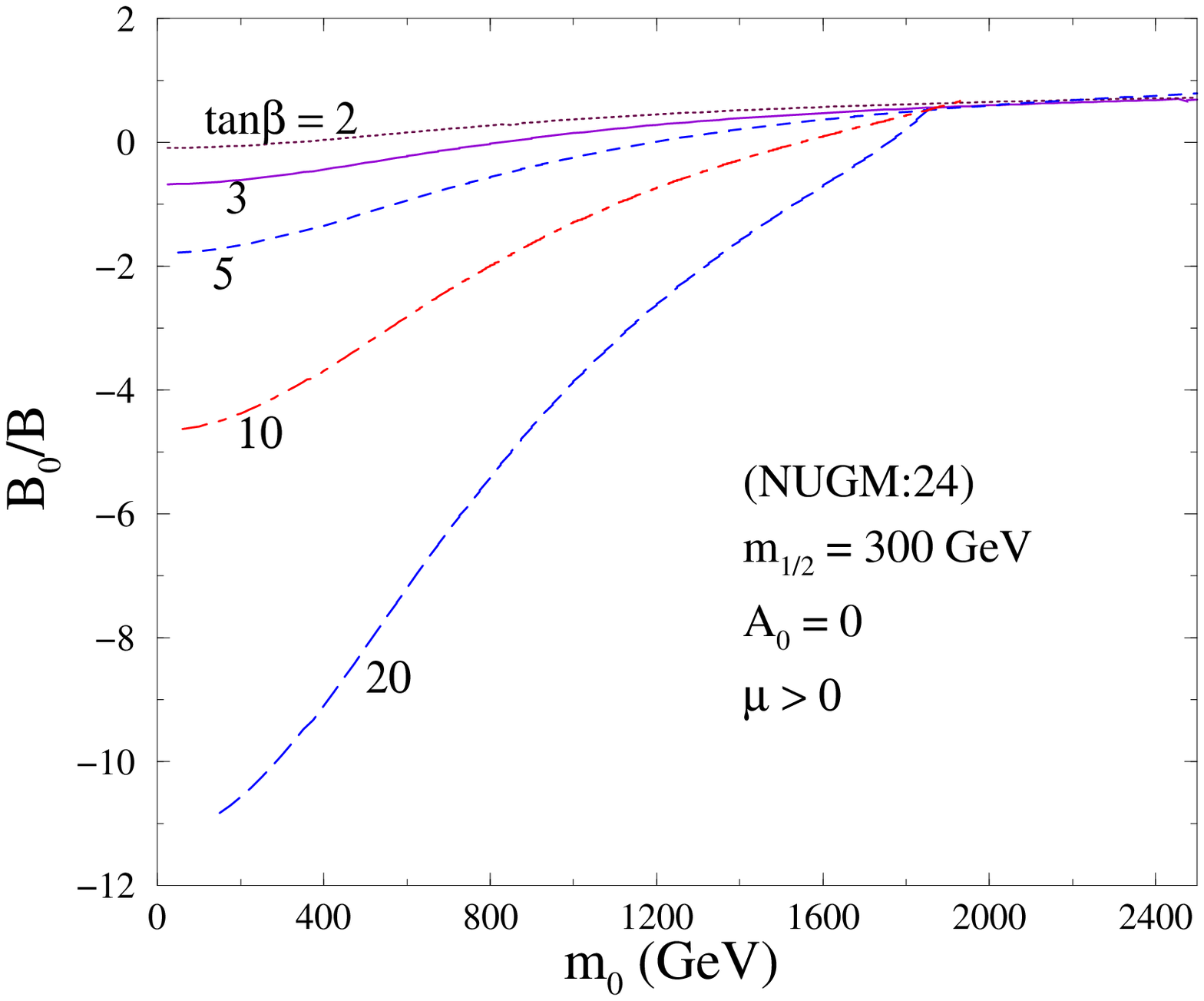}
\hspace*{0.5in}
\mygraph{nonBgraphsD}{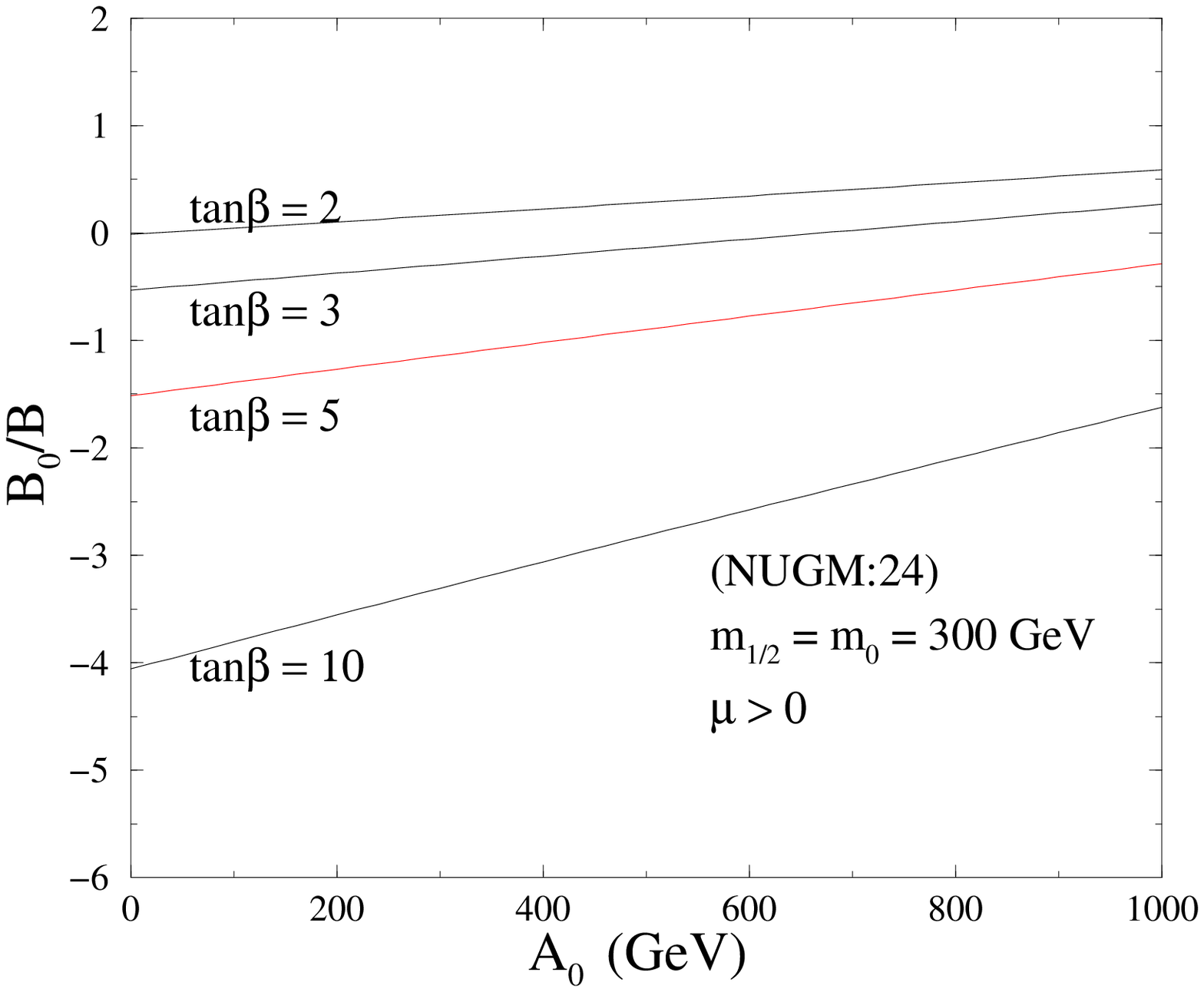}
\caption{\em As in Fig.\protect\ref{sugBgraphs}, but for 
NUGM:24 instead.}                       
\label{nonBgraphs} 
\end{figure} 

We now repeat the analysis for the case of 
NUGM:24 choosing 
$A_0=0$ as before. However, since the sign
of the electroweak gaugino mass parameters are now reversed, the
gaugino contribution to Eq.\ref{EqnBrge} would now enhance the
trilinear contribution instead of cancelling it.  
And since the sign inversion affects 
only the sub-dominant contributions to the evolution of 
$A_{t, b}$, the latter remain close to their mSUGRA values with the 
result that the total  trilinear contribution to $d B / d t$  suffers
only a  small relative change.
The result is then a monotonic decrease of $B_0$ with an
increase in $\mhalf$, and hence, in an appreciably large amount of evolution
(Fig.\ref{nonBgraphsA}).

A further consequence is that the ratio $B_0/B$ too is monotonic in
$\mhalf$ (Fig.\ref{nonBgraphsB}).  The slope though decreases with
$\mhalf$, leading to a flat behavior for moderately large $\mhalf$
values. This can be understood by realizing that, apart from $B$ being
approximately linear in $\mhalf$ $\Delta B$ too is approximately
linear especially for large $\mhalf$.  While the steep
slope for small $\mhalf$ might seem intriguing given the almost 
linear behavior of both $B$ and $B_0$ in Fig.\ref{nonBgraphsA}, 
it should be noted that $B$ is very small for such $\mhalf$ and consequently
any departure from linearity would be magnified in the ratio.
That the slopes at small $\mhalf$ values grow with 
$\tanb$ is understandable too, as for larger $\tanb$, the trilinear 
term contributions to $\Delta B$ assume greater significance. 

The abrupt ending of the curves, especially for larger $\tanb$ values
might seem curious.  However, note that $A_\tau$ is appreciably larger
in NUGM:24 than in mSUGRA (see Sec.\ref{AnalysisSubSec1}). This leads
to a rapid suppression of $m_{{\tilde \tau}_1}$, the mass of the
lighter stau. While the latter also sees an enhancement on account of
the $SU(2)$ gaugino mass being significantly larger in NUGM:24 in
comparison that within mSUGRA for an identical value of $\mhalf$, this
effect is sub-dominant.  Consequently, for such parameter values, the
lighter stau would have a mass smaller than the lightest of the
neutralinos thereby becoming the lightest supersymmetric
particle. Since this is phenomenologically unacceptable, such regions
of the parameter space have to be discarded. Note though that the
extent of the allowed parameter range in the $\mhalf$--$\tanb$ plane
does depend on the value of $m_0$.

Fig.\ref{nonBgraphsC} displays 
$B_0/B$ for different values of $\tanb$
as $m_0$ is varied. As discussed before, $B$ increases with 
increase of $m_0$ and diminishes with increasing $\tan\beta$. 
For most of the region (except when $m_0$ is large and $\tan\beta$ 
quite small) the ratio can be large and negative because of  
a large degree of evolution of $B$ in NUGM:24. 
For larger $\tanb$, $B$ 
itself is much smaller. Hence a large evolution results into a large
negative $B_0$. On the other hand, a larger 
value for  $m_0$ pushes $B$ higher and $B_0$ 
would then be dragged down to a value near zero. 
Additionally, we like to  
clarify that the larger $\tan\beta$ curves really 
end near 2 TeV or so in Fig.\ref{nonBgraphsC} because 
of the REWSB requirement. This is unlike the smaller $\tan\beta$ 
contours that span the entire $m_0$ range displayed. 

As for the dependence on $A_0$ (see Fig.\ref{nonBgraphsD}), 
the relationship is once again 
linear, as predicted by Eq.\ref{EqnBzVsB0}, 
for either of the two models under discussion.

\subsection{Evolution of CP violating phases} 
\label{ResultsSubSec2}
\begin{figure}[!h]
% \vspace*{-1.0in}                                 
\mygraph{compBgraphsug}{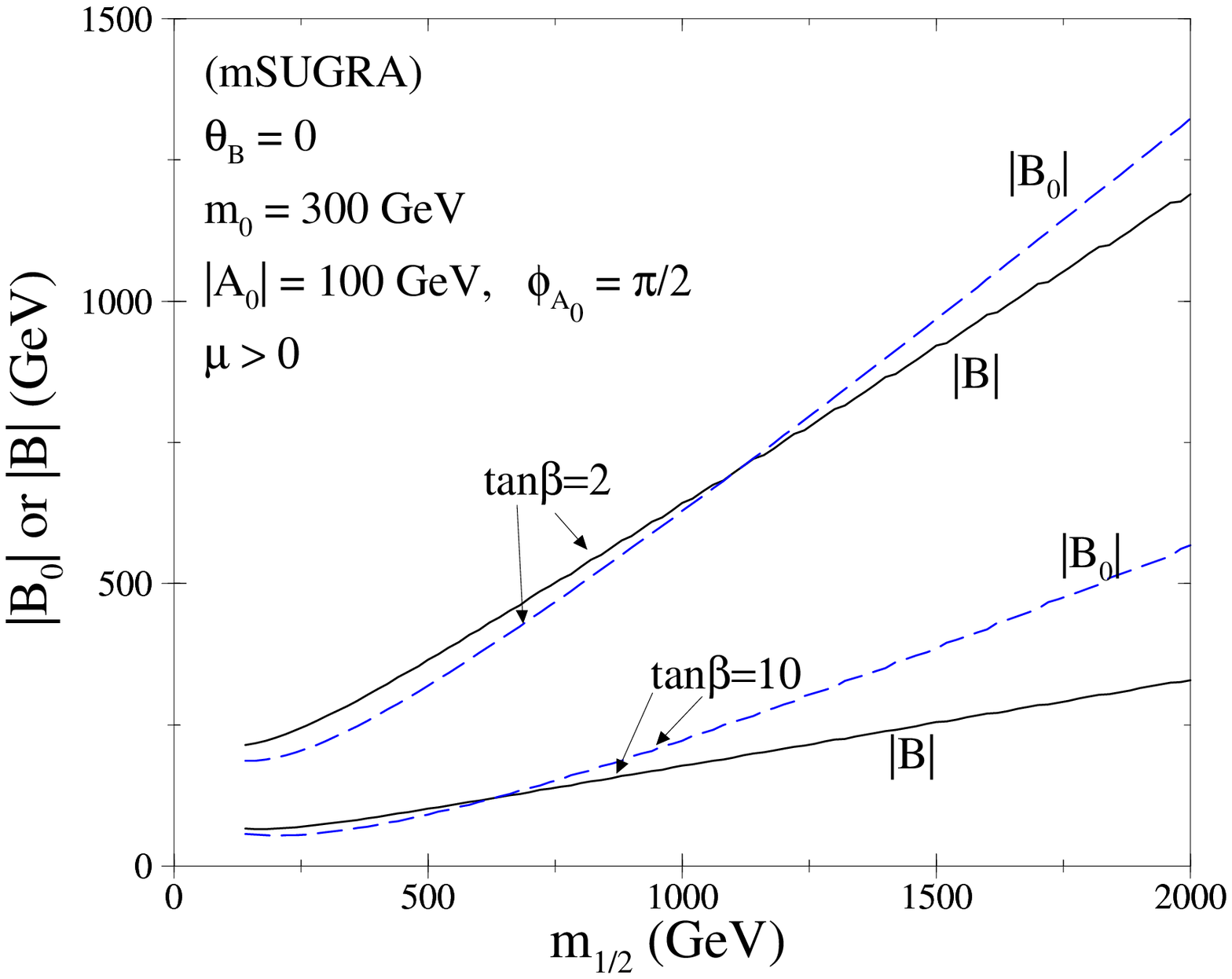}
\hspace*{0.5in}
\mygraph{compBgraphnon}{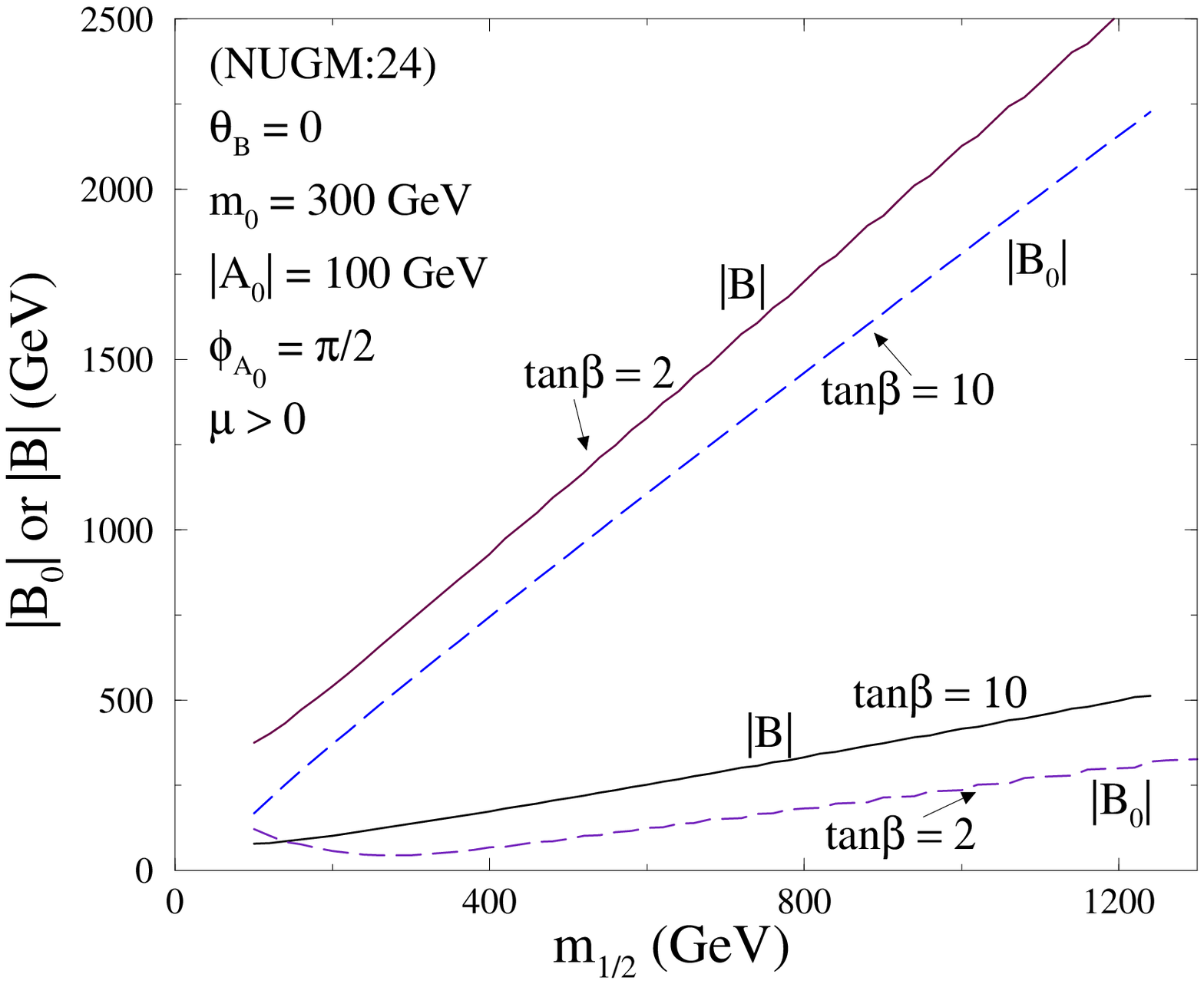}
\caption{\em $B$ and $B_0$ vs $\mhalf$ for the displayed 
parameters with non-zero $\phi_{A_0}$ in mSUGRA and in NUGM:24.}     
\label{compBgraph} 
\end{figure}

Having analyzed the simple case of $\theta_B = \phi_{A_0} = 0$, we may
now consider the effect of phases. To start with, we continue to
maintain $\theta_B = 0$, but now consider $ \phi_{A_0} = \pi / 2$, or,
in other words, a maximal phase in the trilinear coupling.  This
choice maximizes the EDM values\cite{EDMmany2}.  To study the generic
features and compare with the results of Sec.\ref{ResultsSubSec1}, we
first choose a relatively small value of $|A_0|$ ($=100$~GeV). Thus,
$\Re(A_i)$ and $\Re(B$) would not be very different from the analysis
of Sec.\ref{ResultsSubSec1} because of the absence of any phase in the
gaugino parts of Eqns.(\ref{EqnBrge} \& \ref{EqnAtrge}) and the
smallness of $|A_0|$. With this choice of inputs, the only
contributions to $d \Im(B)/ d t$ or $d \Im(A_i) / dt$ arise from
$\Im(A_i)$ themselves, and hence there is no occasion for
cancellations/enhancements unlike in the case for the real parts.  
In addition, the effect of
$\phi_{A_0}$ on $|B_0|$ would be limited even for maximal $\phi_{A_0}$
unless $|A_0|$ is quite large.  
This is reflected by Figs.\ref{compBgraph}, wherein we
display the variation of both $|B|$ and $|B_0|$ with $\mhalf$ for
either model.  The results are seen to be consistent with the no-phase
cases of Fig.\ref{sugBgraphsA} and Fig.\ref{nonBgraphsA}.

\begin{figure}[!h]           
% \vspace*{-1.0in}                                 
%\subfigure[Variation of B-parameters in mSUGRA]{                       
\mygraph{phaseAllA}{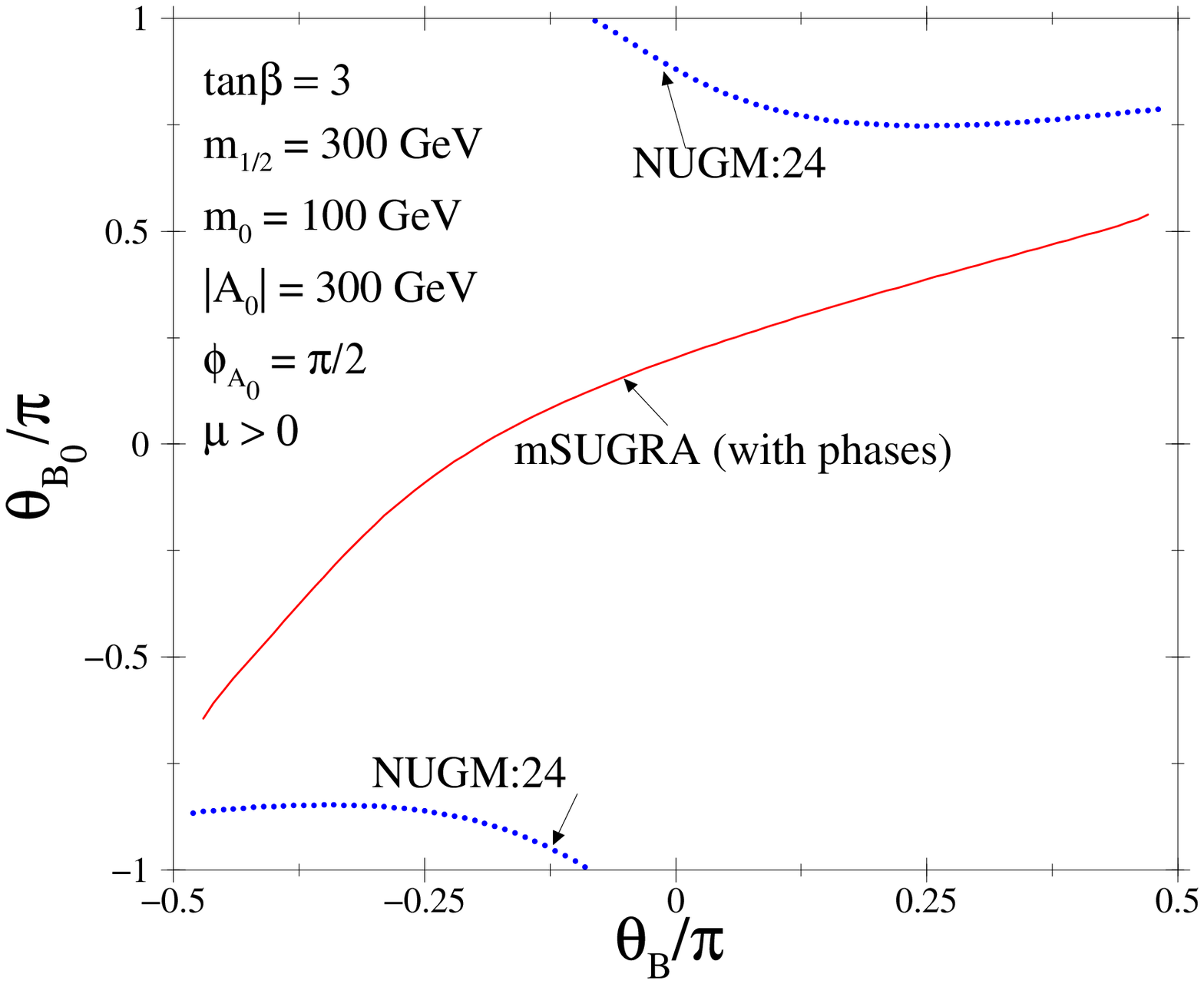}
\hspace*{0.5in}
\mygraph{phaseAllB}{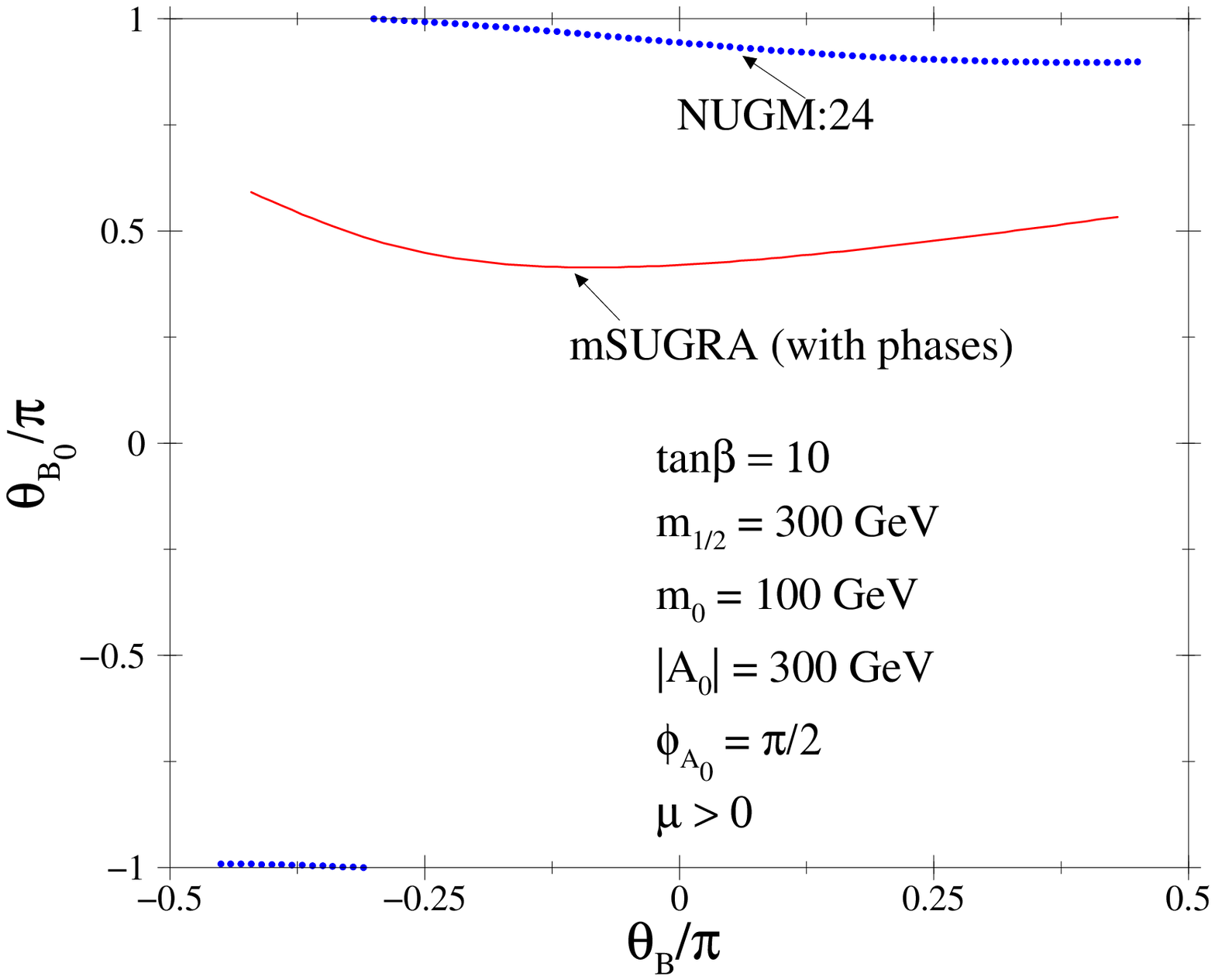}
\caption{\em $\thetab0$ vs $\theta_B$ for $\tan\beta=3$~and 10 with other
parameters are as shown for mSUGRA and NUGM:24 scenarios.  $\thetab0$ 
(defined to lie in the range $[-\pi,\pi]$) is seen to be larger for NUGM:24.  
}  
\label{phaseAll} 
\end{figure}

We now invoke a non-zero $\theta_B$ and analyze
the resulting evolution of the same from the electroweak scale 
to the GUT-scale.  In Figs.\ref{phaseAll}, we display this for
both mSUGRA and NUGM:24, and in each case for two values of $\tanb$,
namely 3 and 10. Again, for illustrative purposes, we choose, for the
other relevant parameters, $m_0=100$~GeV, $\mhalf=300$~GeV and
$|A_0|=300$~GeV with 
$\phi_{A_0}=\pi/2$.  Although the constraints from
the EDM measurements restrict $|\theta_B|$ to very small values
($\lsim {\cal O}(10^{-2})$), we display the functional dependence for
a wider range of $\theta_B$.  The apparent discontinuities for the
NUGM:24 curves are not physical and have only been occasioned by the
choice for the domain of $\thetab0$, namely $[-\pi, \pi]$.  
Clearly, the amount of phase evolution in NUGM:24 is seen to be 
higher than that in mSUGRA. 

\begin{figure}[!h]
\mygraph{PhiGraphA}{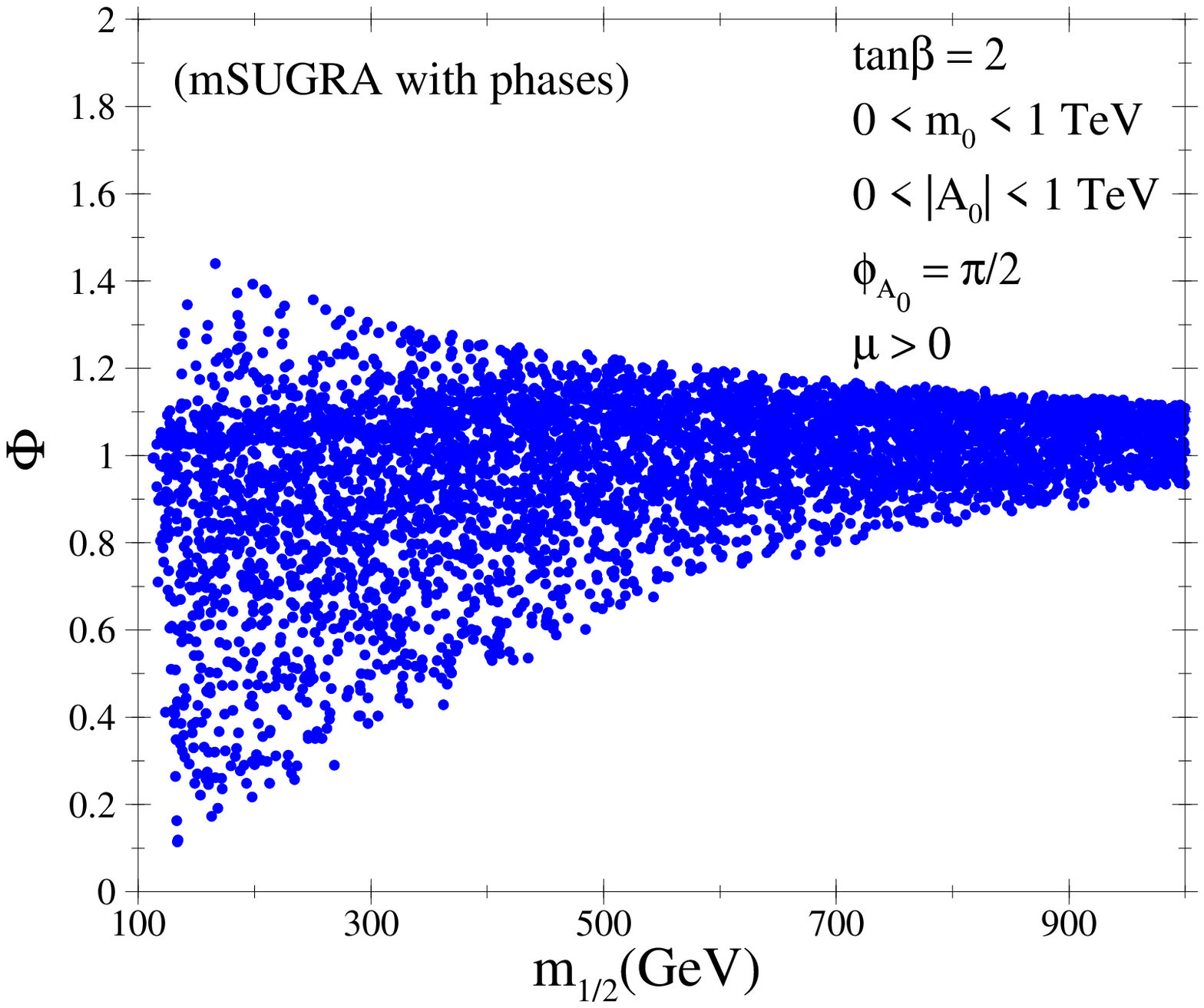}
\hspace*{0.5in}                     
\mygraph{PhiGraphB}{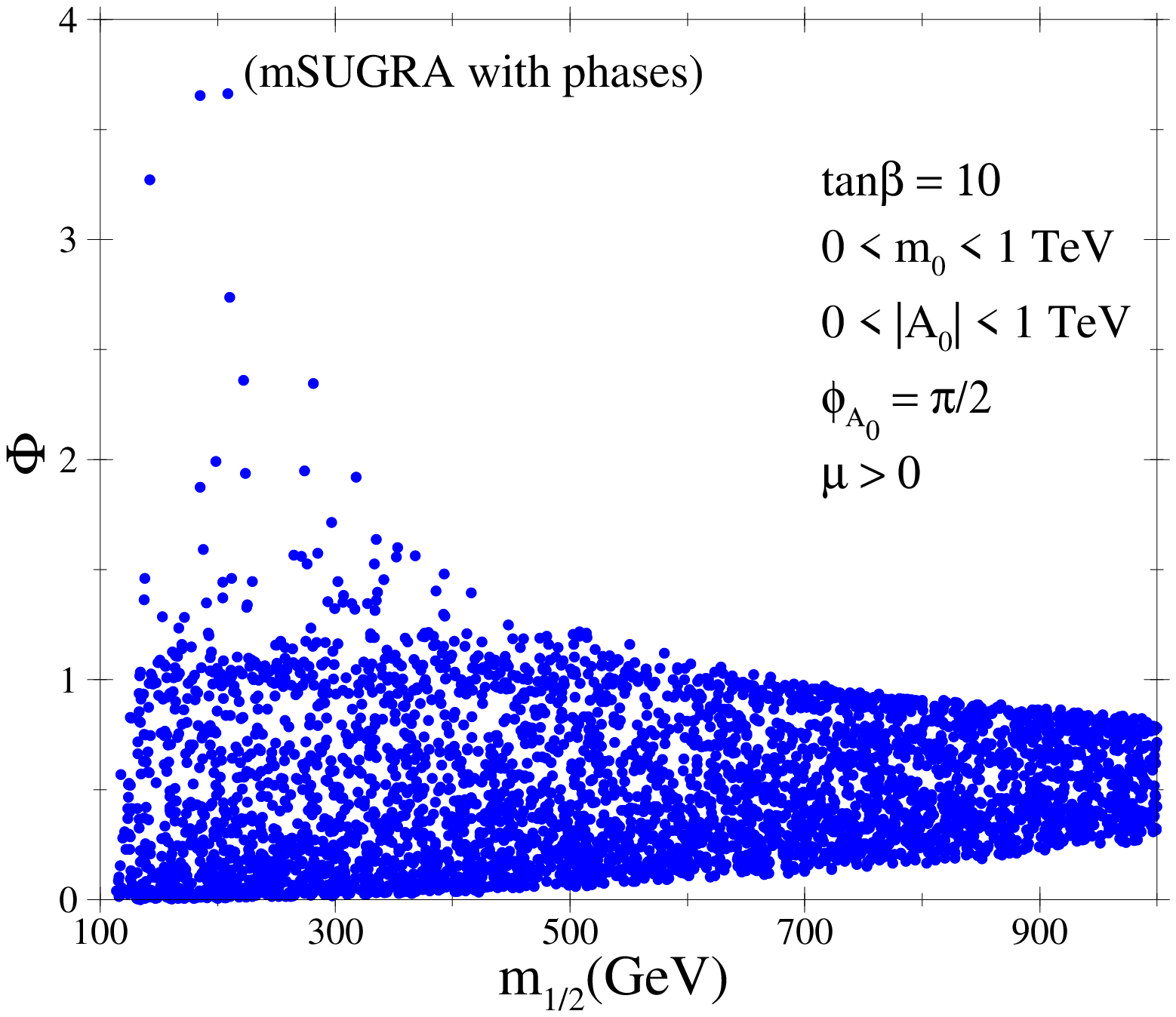} 

\mygraph{PhiGraphC}{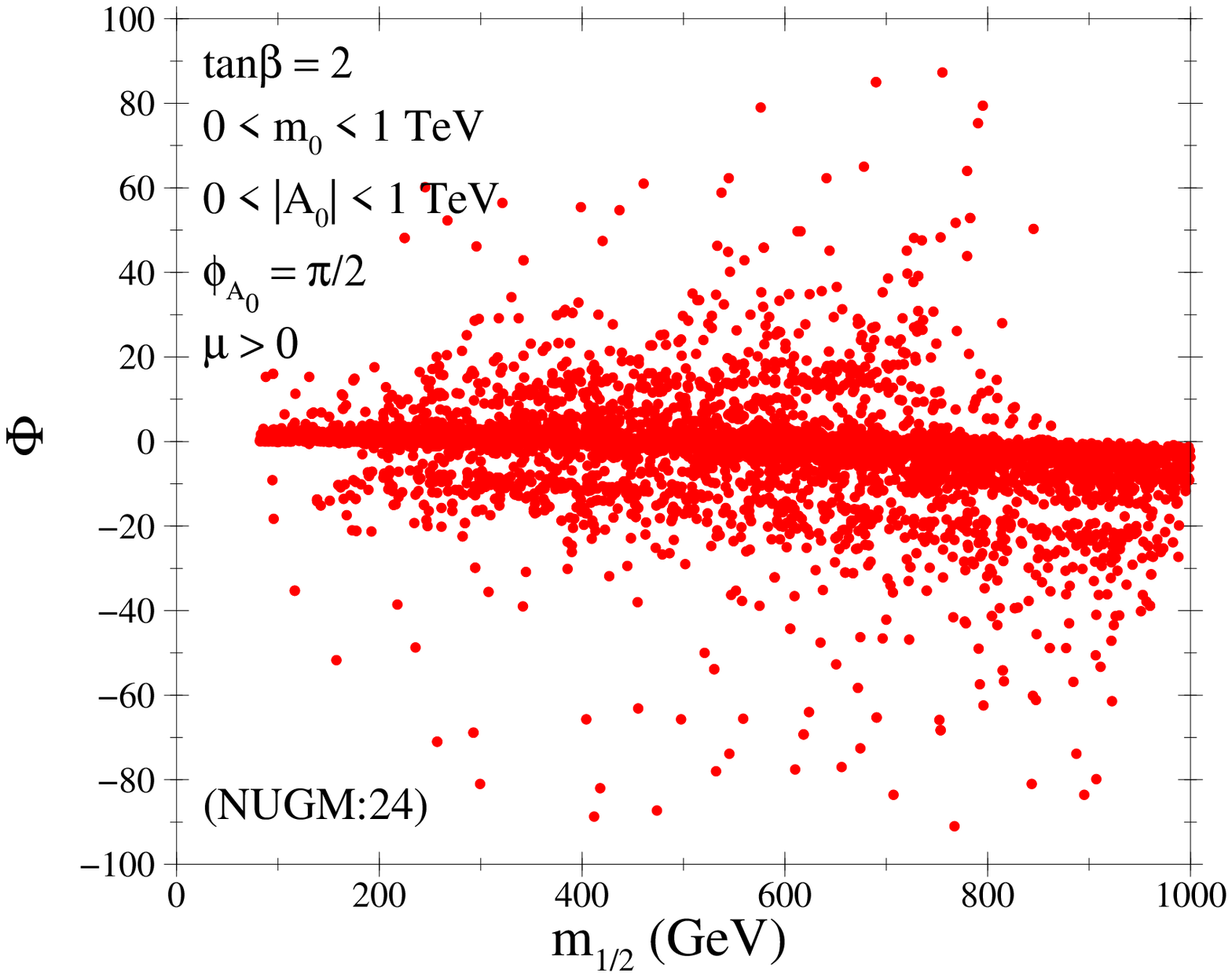}
\hspace*{0.5in}                       
\mygraph{PhiGraphD}{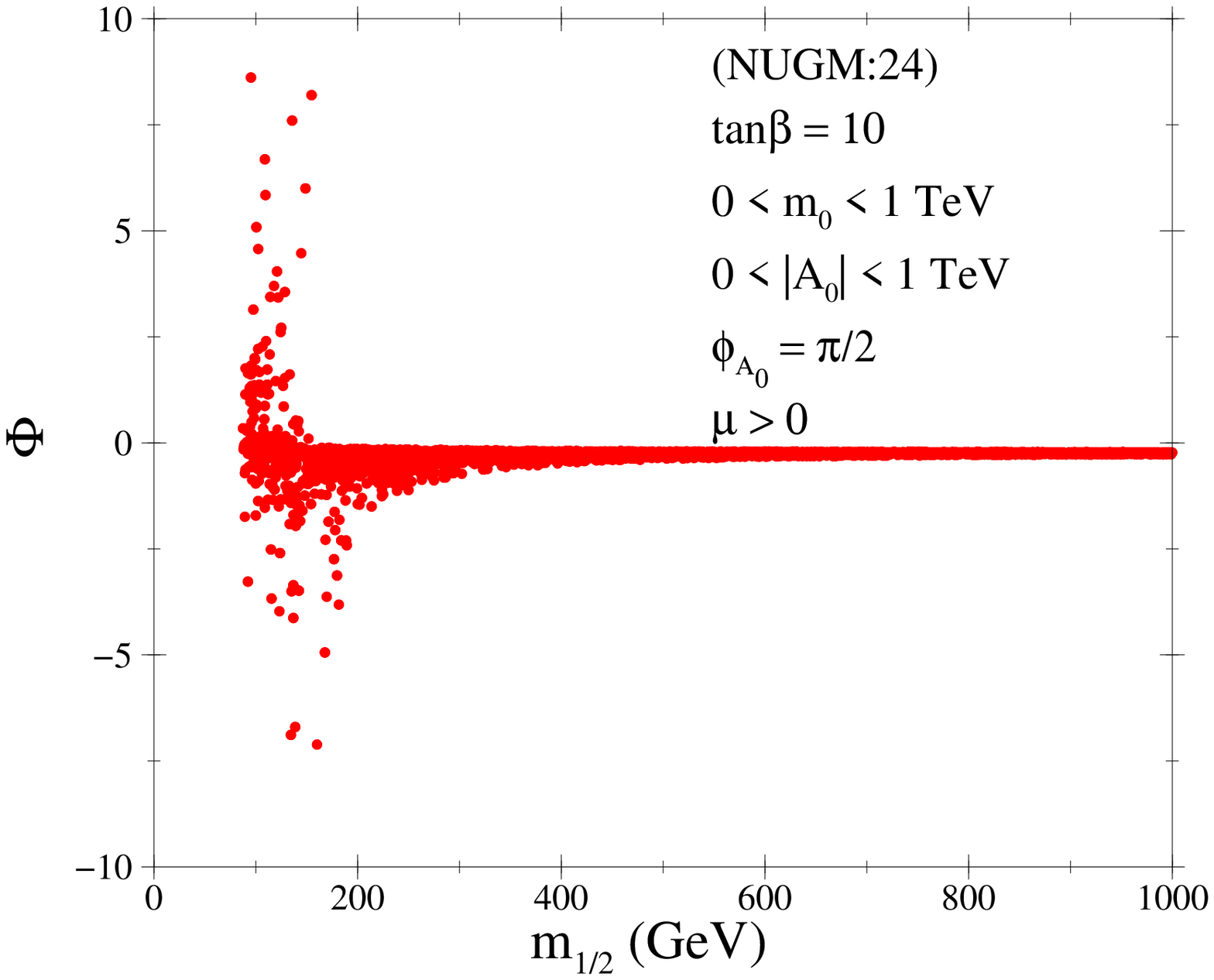}
\caption{\em 
$\Phi$ vs $\mhalf$ for mSUGRA and NUGM:24 for $\tan\beta=2$ and $10$, 
when $m_0$ and $|A_0|$ are scanned up to 1~TeV for $\phi_{A_0}=\pi/2$. }
\label{PhiGraphAll} 
\end{figure}

Having established that the degree of fine tuning could, in principle, be 
smaller in the NUGM:24 case, we now perform a 
scan of the parameter space for both mSUGRA and NUGM:24 so as to 
quantify the extent of this reduction. In each case, we consider 
two different values of $\tanb$ ($=2, 10$) while maintaining 
$\phi_{A_0} = \pi/2$ so as to maximize the EDM values. 
Allowing $m_0$, $\mhalf$ and $|A_0|$ to vary up to 1 TeV (with the lower 
end set in accordance with the current limits on super-particle masses), 
we show, in Figs.\ref{PhiGraphAll}, the scatter plots in the $\Phi$--$\mhalf$ 
plane. It is interesting to note that, for low to moderate values of
$\tan\beta$, the measure $\Phi$ rarely  becomes negative 
in the mSUGRA case, whereas in the non-universal scenario it is more evenly 
distributed. 

While $|\Phi|$ does tend to concentrate around zero 
(Fig.\ref{PhiGraphC}), 
note that, for small $\tanb$, the NUGM:24 case 
does have a significantly dense
distribution up to $|\Phi| \sim 20$ and values as large as 
$|\Phi| \sim 100$ are also obtained, albeit with a 
reduced frequency. 
In contrast, the mSUGRA case barely 
registers a presence even for $\Phi \sim 1.5$ (Fig.\ref{PhiGraphA}). 
Thus, in going from mSUGRA to NUGM:24, the fine-tuning can be reduced 
by a factor as large as $\sim 70$. For the $\tanb =10 $ case though, the 
improvement is much more moderate. 
As Fig.\ref{PhiGraphB} shows, the mSUGRA scatter 
reaches up to $\Phi \sim 3.5$, whereas the non-universal scenario admits 
$|\Phi| \sim 10$ (Fig.\ref{PhiGraphD}), or, in other words, a reduction of the
maximal fine tuning by a factor of $\sim 3$. More important, though, 
is that the density of points at higher $\Phi$ is much larger in the 
NUGM:24 case than for mSUGRA. In other words, it is far more likely to 
have a less fine-tuned point in the parameter space for NUGM:24.

\begin{figure}[t]
\vspace*{-0.2in} 
\mygraph{slopeContoursA}{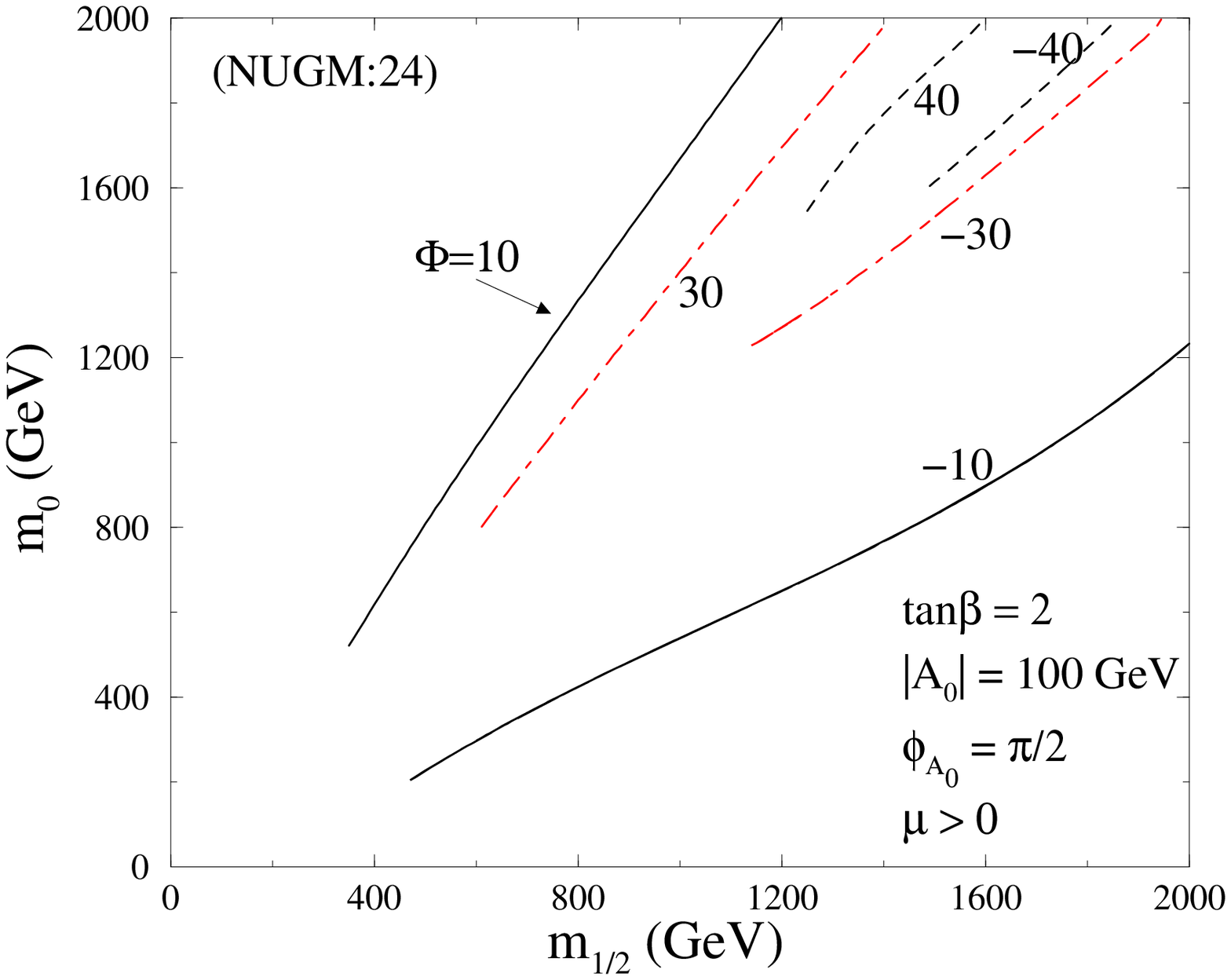}
\hspace*{0.5in}
\mygraph{slopeContoursB}{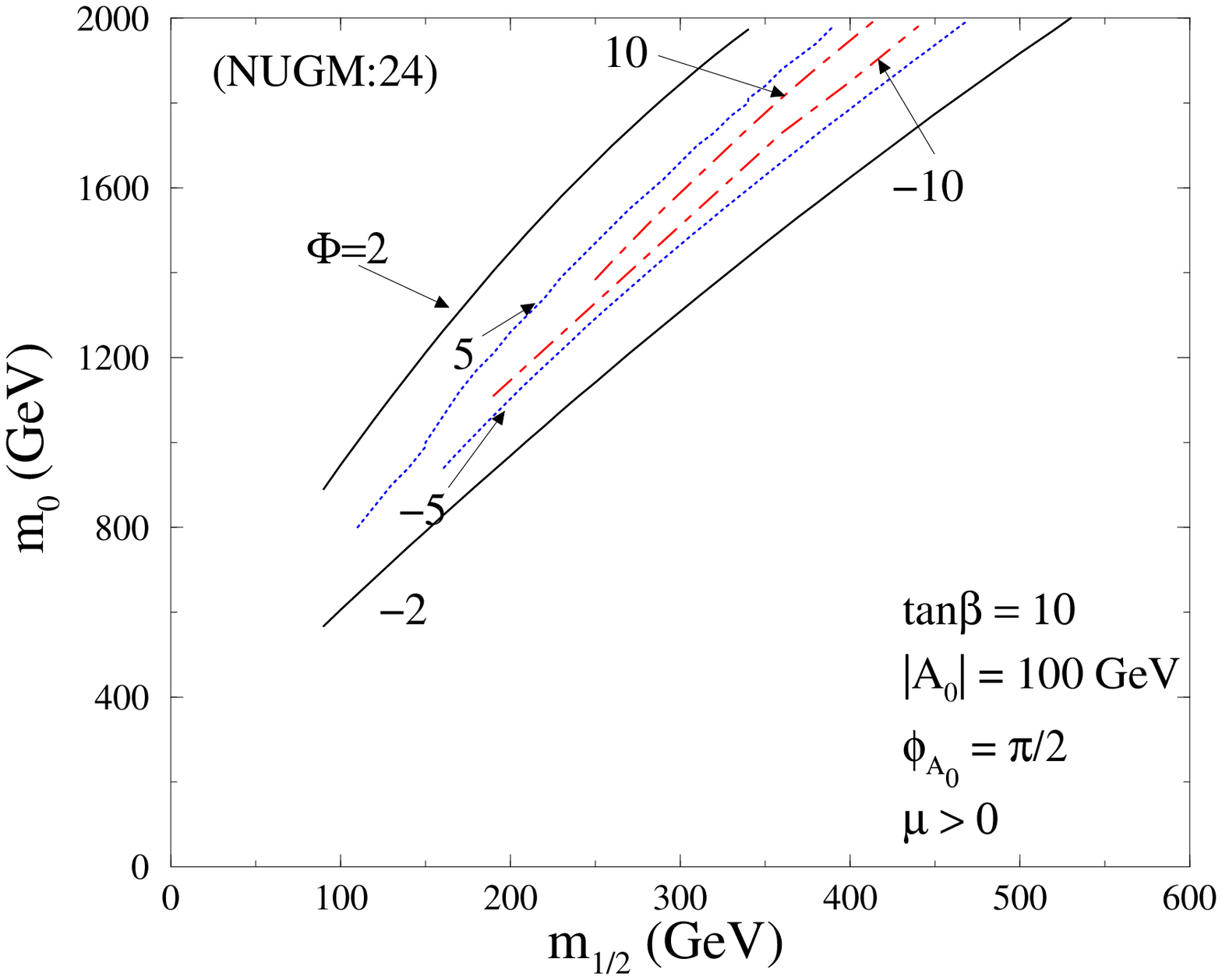}
\caption{\em Contours of $\Phi$ in $\mhalf-m_0$ plane for
$\tan\beta=2$ and $10$ in NUGM:24.  A larger $|\Phi|$ corresponds to a
lesser degree of phase sensitivity.  Switching of the sign of $\Phi$
in some region of parameter space is associated with $\thetab0$
crossing $\pi/2$.  Here $|A_0|=100 \rmGeV$ and $\phi_{A_0}=\pi/2$.
Smaller $\tan\beta$ cases have larger values for $|\Phi|$.}
\label{slopeContours} 
\end{figure}

Concentrating on NUGM:24, we present, in Fig.\ref{slopeContours}, 
contour plots for $\Phi$ in the $m_0-\mhalf$ plane for two different values 
of $\tanb$. Note that the limits on $m_0$ and
$\mhalf$ are $2$ TeV, higher than what was chosen for 
Fig.\ref{PhiGraphAll}. Once again, $|A_0|$ is fixed at $100 \rmGeV$ with
$\phi_{A_0}=\pi/2$.  A comparison of the two
plots clearly reinforces our earlier result that 
the fine-tuning is less severe for low $\tanb$. 
Furthermore, the values of $m_0$ and $\mhalf$ leading 
to a particular $\Phi$ are highly correlated.  
Note that both signs for $\Phi$ are possible. The region where 
$\Phi$ changes sign is associated with a parameter point where
$\thetab0$ is $\sim \pi/2$.
To summarize, the results displayed in
Fig.\ref{PhiGraphAll} and Fig.\ref{slopeContours} show that it is
indeed possible to obtain a surprisingly large amount of reduction of phase
sensitivity even for relatively small sparticle masses.

\begin{figure}[!h]
\mygraph{sugSlopeSmallB0}{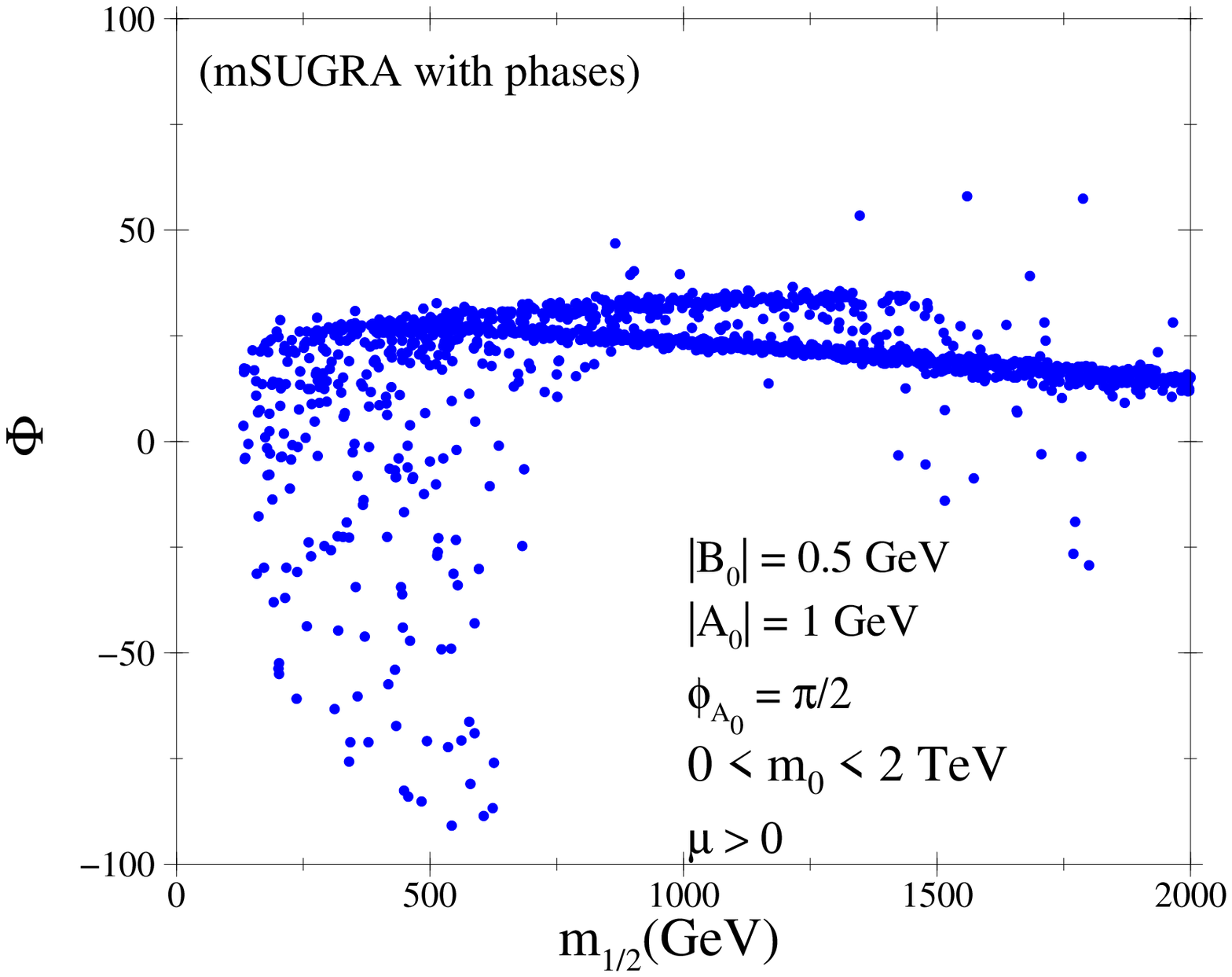}
\hspace*{0.5in} 
\mygraph{nonSlopeSmallB0}{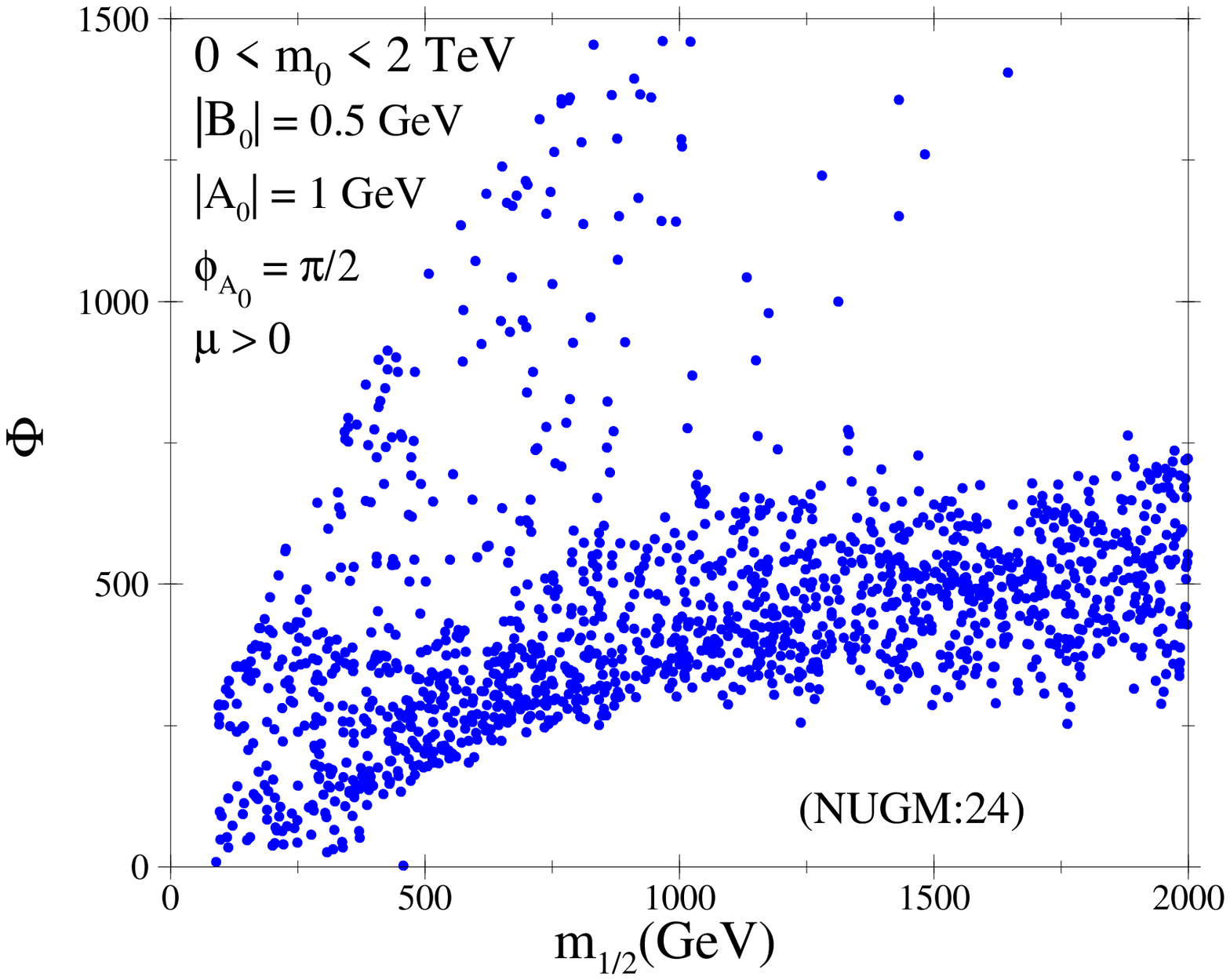}

\mygraph{sugTanSlopeSmallB0}{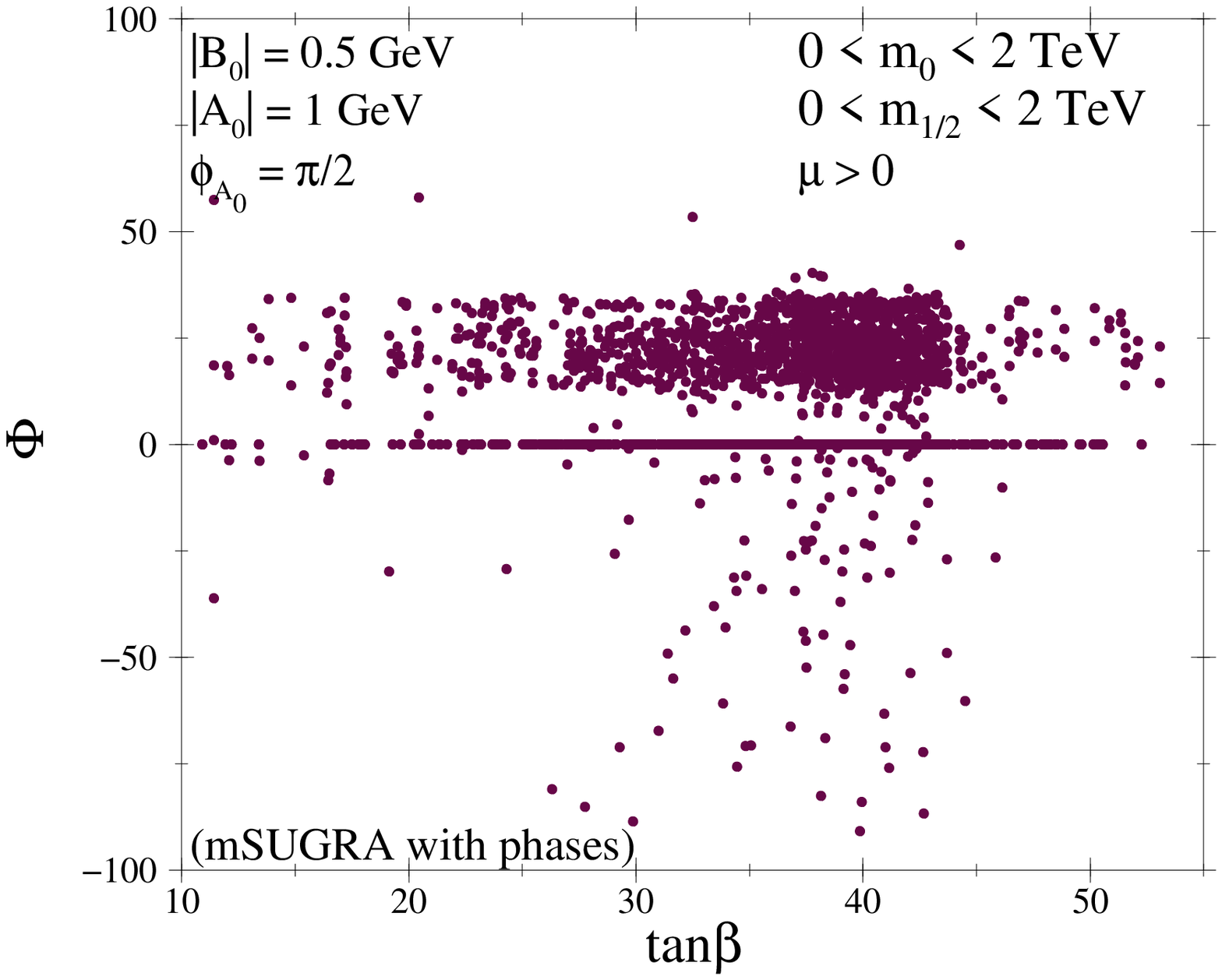}
\hspace*{0.5in}
\mygraph{nonTanSlopeSmallB0}{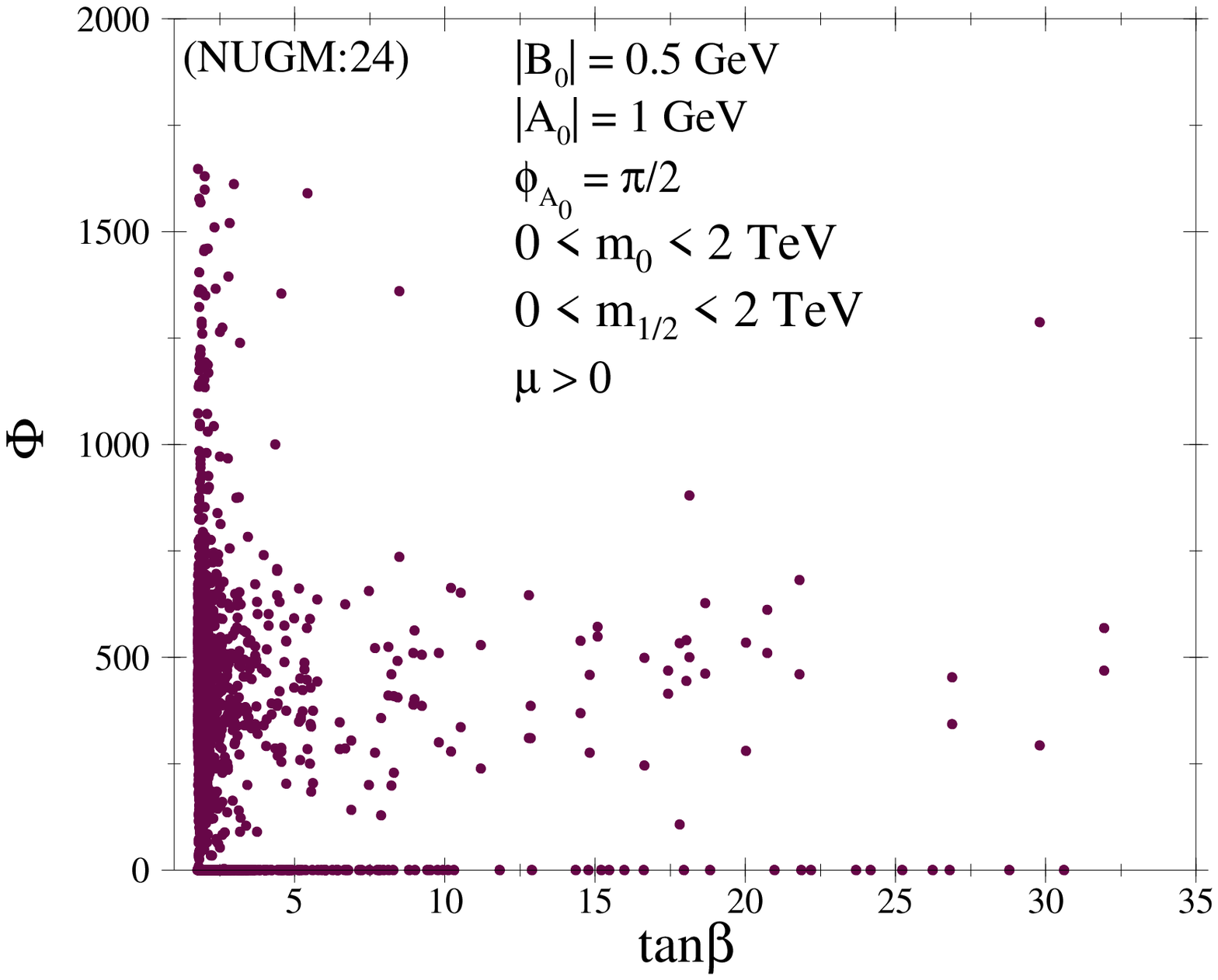}
\caption{\em {\em a)} and {\em b)}: 
Scatter plots corresponding to large values of $|\Phi|$
for mSUGRA and NUGM:24 cases.  
A smaller value for $|B_0|$ as well as smaller $|A_0|$ enhance
$|\Phi|$ both in mSUGRA and NUGM:24.  However, NUGM:24 is associated
with much larger values for $|\Phi|$ as compared to mSUGRA. 
{\em c)} and {\em d)}: Displays of the associated
values of $\tan\beta$ vs $\Phi$.  $\tan\beta$ is small (2 to 5) for
NUGM:24, whereas the same for mSUGRA is large (20 to 45).
}
\label{slopeSmallB0}
\end{figure}

We now explore, in detail, the range of 
$\tan\beta$ that is associated with very low level of phase
sensitivity or, in other words, a very large $|\Phi|$.
As has been argued earlier, $|B|$ itself strongly
depends on $\tan\beta$. Moreover, $\Delta B$, and thereby $B_0$ too, 
has a nontrivial dependence on $\tan\beta$. Thus it is understandable that 
a very large $|\Phi|$ would indeed prominently highlight such a dependence.  
Rather than attempting a full, but very computing-intensive,
scan over the entire parameter space, we choose to restrict 
ourselves to the subset of the parameter space that
would naturally produce very large values for $|\Phi|$, namely the
region with small $|B_0|$ and small $|A_0|$.  
Hence we adopt a
framework with given values for $|B_0|$ instead of $\tan\beta$. The
requirement of REWSB determines $\tanb$ once $B_0$, $m_0$, $\mhalf$
and $A_0$ are fixed.  Note however, that the point $A_0 = B_0 = 0$
would imply the absence of any SUSY CP phase {\em at all scales}.  
Thus, it is not surprising to obtain very large values
of $|\Phi|$ in this scenario.  However, in this part of our work the
focus is simply to study, the effect of $\tan\beta$ on $\Phi$ in
detail, more importantly for large $|\Phi|$ values.  
To quantify our study of this issue, we choose small representative
values viz. $|B_0| = 0.5 \rmGeV$ and $|A_0|=1 \rmGeV$, along with
$\phi_{A_0}=\pi/2$ so as to maximize the EDM contributions as before.
In Figs.\ref{slopeSmallB0}, we present various scatter plots for
$\Phi$ as $m_0$ and $\mhalf$ are varied over a wide range (0 to 2
TeV). Note that the results of this analysis have a significant
dependence on $|A_0|$.  For example, increasing $|A_0|$ to $100$ GeV
may reduce $\Phi$ by a factor of 10 to 20.  As
Fig.\ref{sugSlopeSmallB0} shows, within mSUGRA, $|\Phi|$ could be as
large as $100$ while most of the points lie 
between $10$ to $25$.  The situation is qualitatively different in
NUGM:24 (Fig.\ref{nonSlopeSmallB0}) where $|\Phi|$ may go up to $1500$
while typically ranging between $200$ to $600$.  Thus, 
NUGM:24 is much better able to accommodate low phase-sensitivity
solutions than do the universal gaugino mass scenarios.

It is curious to note that, unlike what Fig.\ref{PhiGraphAll}
suggested, $\Phi$ could assume negative values within mSUGRA (see
Fig.\ref{sugSlopeSmallB0}). This prompts us present a scatter plot of
$\Phi$ against the derived quantity $\tanb$.  As
Fig.\ref{sugTanSlopeSmallB0} shows, mSUGRA admits negative $\Phi$ only
for large $\tanb$. In fact, even for the positive branch, large values
of $|\Phi|$ are typically concentrated in the large $\tan\beta$ (20 to
45) region.  In contrast, for NUGM:24, $\Phi$ assumes larger values
typically for low $\tan\beta$ values (2 to 5). It should be remembered
in this context that, within NUGM:24, the large $\tanb$ domain is 
significantly restricted from considerations of the LSP (see
Sec.\ref{ResultsSubSec1}).  That the favored range for
$\tan\beta$ is different in the two scenarios is attributable to the
interplay between the cancellations/enhancements in the RGE evolution
of $B$ on the one hand and the requirement of REWSB on the other.

Finally, we comment on the case of $\mu<0$.  
It turns out that for this branch of $\mu$ and $\phi_{A_0}=\pi/2$, 
one has $|B_0|>|B|$ for almost all the parameter space of NUGM:24.
As a result one finds no advantage toward 
reducing the phase sensitivity.   

\section{Conclusion}
\label{secConclusion}
As is well known, the experimental upper bounds on the electric dipole
moments of the neutron and the electron impose strong constraints on
any source of $CP$ violation in supersymmetric models, in 
particular on the weak scale phase parameters.
For example, in the minimal
supergravity model, $\theta_B$, the phase of the bilinear Higgs
coupling parameter is constrained to be typically smaller than $0.01$,
with only some very limited regions (such as the focus point scenario)
in the parameter space admitting slightly larger
($\lsim 0.1$) values.  This, however, implies a severe fine-tuning
condition for $\thetab0$, the value of the same phase parameter at
the unification scale. In turn, $\phi_{A_0}$, the phase of the
trilinear coupling parameter is also severely fine-tuned. This has
been a longstanding problem with mSUGRA-like scenarios.

To quantify this problem, we define a {\it phase naturalness}
measure $\Phi$  as the ratio
of the spread of the phase $\thetab0$ at the unification scale that is
consonant with the spread $\theta_B$ allowed, at the electroweak
scale, by the electric dipole moment constraints  A larger $\Phi$
would imply a lower degree of phase sensitivity.  One finds that,
unless $\tan\beta$ is very large, $\Phi$ may be approximated to 
$ B / B_0$ for much of the parameter space.

In this analysis, we have demonstrated that models admitting a large
RG evolution of the bilinear Higgs coupling could be interesting in the context
of a reduction in the fine-tuning of phases.  In particular, we choose
a supergravity-inspired scenario wherein non-universal gaugino masses
arise from a gauge kinetic energy function $f_{\alpha \beta}$
transforming as a particular 
non-singlet representation of $SU(5)$ (NUGM:24 of
Table~\ref{tabrelativeweights}).  As in the mSUGRA (singlet $f_{\alpha
\beta}$) case, this representation, considered in isolation,
introduces no additional phase for the gaugino masses.

Studying the nature of the evolution of $B$ to understand the
correspondence with phase-sensitivity, we identify the large
cancellations in the RGE for $B$ as being primarily responsible for
the high degree of fine-tuning within mSUGRA.  In the NUGM:24, on
the other hand, the said cancellations are replaced by enhancements
(on account of the reversal in the sign of the gaugino mass terms) and
this translates into a reduction of the above-mentioned fine-tuning.
In fact, $\Phi$ can be significantly increased in NUGM:24 (by a factor
of 10 to 20) with respect to comparable mSUGRA type of models. The 
said improvement is typically more pronounced for 
small $\tan\beta$ values.

A particularly interesting result is the identification of extended
regions in the NUGM:24 parameter space which admit a low degree of
phase-sensitivity even for relatively small super-particle masses. This
feature is absent in mSUGRA as well as in most other models with
high scale inputs for SUSY breaking. 

We further explored the dependence of our results, on $\tan\beta$, 
by specifically concentrating on the parameter space 
corresponding to very large $\Phi$ (or very small phase sensitivity) so as to 
compare the two models.  Naturally, this occurs close to 
vanishing $A_0$ and $B_0$ values.  
We adopt a scheme where $B_0$ itself is given as an input parameter 
instead of $\tan\beta$, given the more direct relationship of 
$B_0$ with $\Phi$.
Our analysis shows that, even here, 
the values of $\Phi$ in NUGM:24 are typically larger by a factor of 
10 to 20 in comparison to those in mSUGRA.
And whereas mSUGRA generically requires large
$\tan\beta$ (20 to 40) for $|\Phi|$ to be large, 
the NUGM:24 scenario prefers a smaller $\tan\beta$ (2 to 5) instead.

Finally, while our analysis has focussed on $SU(5)$ 
as the GUT gauge group,  similar considerations hold for 
$SO(10)$ as well. A suitable non-singlet representation resulting in a similar 
gaugino-mass pattern as in NUGM:24 would also produce such
a reduction of phase sensitivity.

\noindent
{\bf Acknowledgments}\\ 
 DC acknowledges financial assistance from the
Department of Science and Technology, India under the Swarnajayanti
Fellowship grant.  DD would like to thank the Council of Scientific
and Industrial Research, Govt. of India for the support received as a
Junior Research Fellow.

\end{document}